\documentclass[11pt]{article}

\usepackage[final]{acl}

\usepackage{times}
\usepackage{latexsym}

\usepackage[T1]{fontenc}

\usepackage[utf8]{inputenc}

\usepackage{microtype}

\usepackage{inconsolata}

\usepackage{graphicx}

\usepackage{amsmath}
\usepackage{amssymb}
\usepackage{cleveref}
\usepackage{xcolor}
\usepackage{bbm}
\usepackage{tipa}
\usepackage{tabularx}
\usepackage{booktabs}
\usepackage{enumitem}

\crefname{section}{\S}{\S\S}
\crefname{appendix}{\S}{\S\S}
\setlist[itemize]{itemsep=0pt,topsep=0pt,parsep=0pt}

\title{[b] $=$ [d] $-$ [t] $+$ [p]: Self-supervised Speech Models Discover \\ Phonological Vector Arithmetic}

\author{
 \textbf{Kwanghee Choi\textsuperscript{1}},
 \textbf{Eunjung Yeo\textsuperscript{1}},
 \textbf{Cheol Jun Cho\textsuperscript{2}},
 \textbf{David Harwath\textsuperscript{1}},
 \textbf{David R. Mortensen\textsuperscript{3}}
\\
 \textsuperscript{1}UT Austin,
 \textsuperscript{2}UC Berkeley,
 \textsuperscript{3}CMU
}

\makeatletter
\newif\ifanon
\ifacl@anonymize
  \anontrue
\else
  \anonfalse
\fi
\makeatother

\makeatletter
\newif\ifanon
\ifacl@anonymize
  \anontrue
\else
  \anonfalse
\fi
\makeatother

\begin{document}
\maketitle
\begin{abstract}
Self-supervised speech models (S3Ms) are known to encode rich phonetic information, yet how this information is structured remains underexplored.
We conduct a comprehensive study across 96 languages to analyze the underlying structure of S3M representations, with particular attention to phonological vectors.
We first show that there exist linear \textit{directions} within the model's representation space that correspond to phonological features.
We further demonstrate that the \textit{scale} of these phonological vectors correlate to the degree of acoustic realization of their corresponding phonological features in a continuous manner.
For example, the difference between [d] and [t] yields a voicing vector: adding this vector to [p] produces [b], while scaling it results in a continuum of voicing.
Together, these findings indicate that S3Ms encode speech using phonologically interpretable and compositional vectors, demonstrating phonological vector arithmetic.
All code and interactive demos are available at
\ifanon
\url{https://anonymous.4open.science/r/phonetic-arithmetic-1886}.
\else
\url{https://github.com/juice500ml/phonetic-arithmetic}.
\fi
\end{abstract}

\section{Introduction}
\begin{figure}[t]
\centering
  \includegraphics[width=\linewidth]{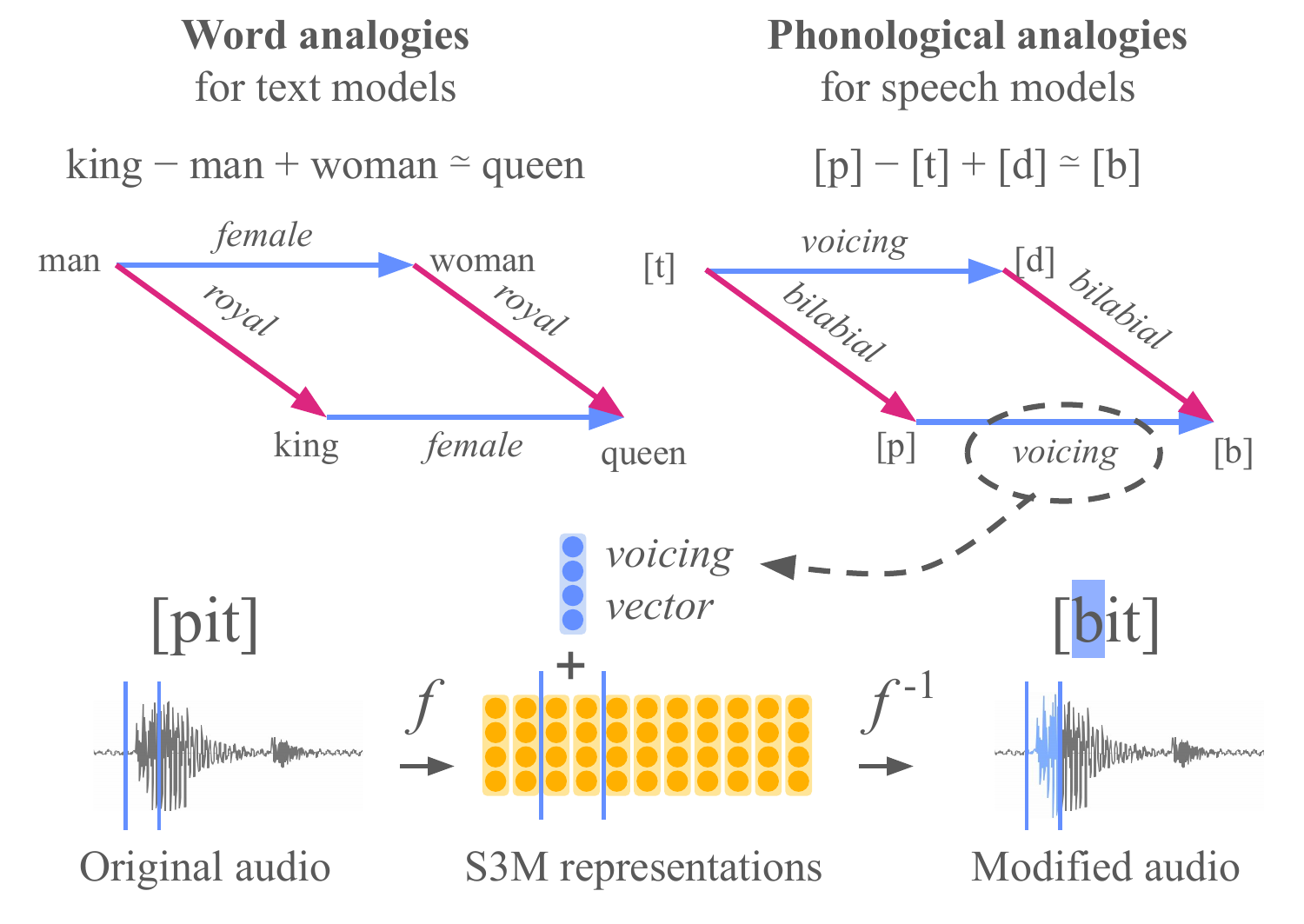}
  \caption{
    Comparing analogies for text and speech. 
    Word representations enable semantic analogies\protect\footnotemark, while speech representations enable phonological analogies.
    Such analogies (\Cref{sec:direction}) can be used to control speech synthesis in a phonologically grounded manner (\Cref{sec:degree}).
  }
  \label{fig:analogy}
\end{figure}

The same year that word2vec \cite{mikolov2013efficient}, a self-supervised method for distributional dense word representations, was introduced, \citet{Mikolov2013LinguisticRI} showed that these representations encode linguistically meaningful relations through vector arithmetic (\cref{fig:analogy}), \textit{i.e.},
\begin{align}
    \mathbf{v}_\text{king} - \mathbf{v}_\text{man} + \mathbf{v}_\text{woman} \simeq \mathbf{v}_\text{queen},
\end{align}
providing a ready explanation for how word2vec represents semantics.\footnote{Many have argued that the analogy test used in \citet{Mikolov2013LinguisticRI}, which differs from ours, can be fragile in practice; see \Cref{ss:fragile} for details.}
In this paper, we ask: Do self-supervised speech models (S3Ms) represent phonology in an analogous compositional manner?


S3Ms, a class of neural networks trained on large quantities of unlabeled speech, have demonstrated strong performance across various downstream tasks, including speech recognition, synthesis, and spoken language understanding \citep{baevski2020wav2vec,hsu2021hubert,chen2022wavlm}.
Consequently, many studies have sought to understand what properties of S3M representations support this performance.

\footnotetext{Word analogies borrowed from \citet{ethayarajh2019towards}.}

Early analyses primarily investigated \textbf{what} information S3M representations encode \citep{pasad2021layer}.
Empirical studies have shown that S3Ms organize speech according to relative distances that reflect acoustic similarity \citep{choi2022opening}, and that these representations form clusters corresponding to phonetic units \citep{wells2022phonetic,choi2025leveraging}.
However, \textbf{how} this information is structured is still underexplored.

Building on prior observations and linguistic intuition, we aim to refine the current understanding of S3Ms. 
In \Cref{sec:direction}, we propose a hypothesis:
\textit{Phonological features are represented linearly within S3M representations, allowing phonological analogies to emerge.}
For example, consider the phone quadruplet [b], [p], [d], and [t].
[b] and [p] form a voiced-voiceless bilabial plosive pair, and [d] and [t] form a voiced-voiceless alveolar plosive pair.\footnote{Voicing refers to the vibration of the vocal folds. Bilabial and alveolar describe place of articulation (POA), with bilabials produced by bringing both lips together and alveolars by placing the tongue tip against the alveolar ridge.}
This yields two symmetric phonological analogies:
\begin{align}
    \text{[b]}:\text{[p]} &= \text{[d]}:\text{[t]} &(\text{voicing)} \\
    \text{[b]}:\text{[d]} &= \text{[p]}:\text{[t]} &(\text{POA)}
\end{align}
These analogies yield the following approximations in each corresponding vector $\mathbf{r}$ extracted from S3Ms (details in \Cref{ss:models}):
\begin{align}
    \mathbf{r}_\text{[b]} &\simeq \mathbf{r}_\text{[p]} + (\mathbf{r}_\text{[d]} - \mathbf{r}_\text{[t]}) &(\text{voicing)} \label{eq:voicing} \\
    &= \mathbf{r}_\text{[d]} + (\mathbf{r}_\text{[p]} - \mathbf{r}_\text{[t]}) &(\text{POA}) \label{eq:place}
\end{align}
which yields two symmetric compositional phonological vectors: a voicing vector $\mathbf{v}_\text{voi.} = \mathbf{r}_\text{[d]} - \mathbf{r}_\text{[t]}$ in \cref{eq:voicing} and a change of POA vector\footnote{We can also name it as alveolar-to-bilabial vector or negative coronality vector.} $\mathbf{v}_\text{POA} = \mathbf{r}_\text{[p]} - \mathbf{r}_\text{[t]}$ in \cref{eq:place}.

To empirically evaluate this hypothesis, we tested 19 phonological features from PanPhon \citep{mortensen2016panphon}.\footnote{Syllabic, sonorant, continuant, delayed release, lateral, nasal, strident, voice, spread glottis, anterior, coronal, distributed, labial, high, low, back, round, tense, and long.}
We find that analogies based on all 19 phonological features consistently hold in S3M representations.

In \cref{sec:degree}, we further explore the scale of each phonological vectors.
In detail, we introduce a scalar $\lambda \in \mathbb{R}$ into \cref{eq:voicing}:
\begin{align}
    \mathbf{r}_\text{[b]} &\simeq \mathbf{r}_\text{[p]} + \lambda \cdot (\mathbf{r}_\text{[d]} - \mathbf{r}_\text{[t]}). \label{eq:weighted-voicing}
\end{align}
We hypothesize that \textit{the scale $\lambda$ will control the acoustic characteristics associated with a phonological vector in a continuous manner}.
For instance, we expect the scale $\lambda$ of the voicing vector $\mathbf{v}_\text{voi.}$ to correspond to the degree of voicing of the segment.

To validate this hypothesis, we train a vocoder $f^{-1}$ to approximate the inverse of S3M $f$:
\begin{align}
    \mathbf{R} &= f(\mathbf{x})\\
    \mathbf{x} &\simeq f^{-1}(\mathbf{R}),\label{eq:vocoder}
\end{align}
where $\mathbf{x}$ is the input speech signal and $\mathbf{R} \in \mathbb{R}^{T'\times F}$ its S3M representation before pooling (details in \cref{ss:pooling}).
We modify the representation $\mathbf{R}$ by adding scaled phonological vectors (\cref{eq:weighted-voicing}) and resynthesize the audio through the vocoder $f^{-1}$ (\cref{eq:vocoder}).

Eight phonological features that can be directly quantified through acoustic measurements on $\mathbf{x}$ were selected: height (high and low), backness, roundness, nasality, sonority, stridency, and voicing.
We observe that acoustic measurements strongly correlate with the scale $\lambda$, for both interpolation ($|\lambda|\leq1$) and extrapolation ($|\lambda|>1$) of phonological vectors.
These results suggest that S3M representations encode phonological features not as purely binary distinctions but as a continuum through specific vector directions and scales.

In summary, our contributions to advancing the understanding of S3M representations are twofold:
\begin{itemize}
    \item \textbf{Direction}: We show that S3M representations exhibit phonological vector arithmetic, \textit{i.e.}, existence of compositional vectors that align with phonological features (\Cref{sec:direction}).
    \item \textbf{Scale}: We show that the scale of phonological vectors corresponds to the degree of their associated phonological features, leading to interpretable control of speech synthesis (\Cref{sec:degree}).
\end{itemize}

\section{Datasets}\label{ss:datasets}
We use two phonetically transcribed and manually segmented datasets, TIMIT and VoxAngeles, to evaluate both English-specific and cross-linguistic generalizability of our findings.

\textbf{TIMIT} \citep{garofolo1993darpa} contains utterances and their phonetic segmentations of 630 English speakers with balanced phonetic and dialectal coverage.
To avoid ambiguities in extracting phonological features, we exclude diphthongs from our analysis. In addition, plosive closures are merged with their subsequent releases.\footnote{Plosives in TIMIT are separately segmented into closure and burst, different to IPA definitions.}

\textbf{VoxAngeles} \citep{chodroff2024voxangeles} provides phonetic segmentations for audio from the UCLA Phonetics Archive from 95 languages across 21 language families,
providing broader phonetic diversity.
As the dataset has no predefined train/test splits, we use the full set.
Because VoxAngeles excludes English, it enables evaluation of whether S3Ms generalize to phones that do not occur in English.

\section{Experiment 1: Direction of Phonological Vectors}\label{sec:direction}
In this section, we test the first hypothesis: whether linear phonological vectors that satisfy phonological analogies exist within S3M representations, by observing their directions (\cref{eq:voicing,eq:place}).

\subsection{Method}
\subsubsection{Creating phonological analogies}\label{ss:quads}
Phonological features capture fundamental properties of speech sounds \citep{trubetzkoy1939grundzuege, jakobson1951preliminaries}, such as voicing, place of articulation (\textit{e.g.}, bilabial, coronal, velar), and manner of articulation (\textit{e.g.}, plosive, fricative, nasal).
To comprehensively analyze such features, we use PanPhon \citep{mortensen2016panphon} to extract discrete phonological features for each phone $p$: $\mathbf{h}_p' = \texttt{PanPhon}(p) \in \{-1, 0, 1\}^{21}$.
PanPhon yields 21 phonological features with values $+$ (present), $0$ (not applicable), or $-$ (absent).
For example, /b/ can be represented as [$+$voice, $+$labial, $-$nasal, $+$anterior, $0$tense, $\cdots$].
To binarize the features, we expand each value $+$, $0$, and $-$ to be $[1, 0]$, $[0, 0]$, and $[0, 1]$, producing $\mathbf{h}_p = \texttt{extend}(\mathbf{h}_p')\in \{0, 1\}^{42}$.

A quadruplet of phones $\mathbf{p}=(p_1, p_2, p_3, p_4)$ yields two symmetric analogies, \textit{e.g.}, \cref{eq:voicing,eq:place}.
We construct the quadruplet set $Q$ by filtering from every possible quadruplets drawn from the phone vocabulary $\mathbf{p} \in V$:\footnote{
$\mathbf{p}\in V$ means that all phones $p_1, p_2, p_3, p_4$ belong to $V$.
}
\begin{align}
    Q = \{\mathbf{p} \in V | \mathbf{h}_{p_1} -\mathbf{h}_{p_2} = \mathbf{h}_{p_3} -\mathbf{h}_{p_4} \},\label{eq:analogies}
\end{align}
such that each quadruplet $\mathbf{p} \in Q$ denotes two symmetric analogies $p_1 : p_2 = p_3 : p_4$ and $p_1 : p_3 = p_2 : p_4$, or, equivalently, $\mathbf{r}_{p_1} \simeq \mathbf{r}_{p_2} + \mathbf{r}_{p_3} - \mathbf{r}_{p_4}$.
Note that the analogies are not required to be minimal pairs and may differ along multiple phonological features, where we did not observe any systematic influence of minimal vs. non-minimal pairs (\cref{ss:layerwise-pfer}).

We obtain $236$ and $468$ quadruplets (hence double the analogies) from the TIMIT test set and the full VoxAngeles dataset, respectively.
We excluded the quadruplets where there is no PanPhon mapping for any one of the phones, or if any of the phones has less than 50 occurrences within the dataset, to ensure reliable cosine similarity estimates, which reduced the number of phones from 47 to 43 for TIMIT, 567 to 57 for VoxAngeles.
No quadruplets were obtained for the ``constricted glottis'' and ``consonantal'' feature for both datasets,\footnote{low, constricted glottis, spread glottis, long, and consonantal in TIMIT, and constricted glottis, lateral, and consonantal in VoxAngeles were missing.} resulting in $21-2=19$ phonological features being tested.

\subsubsection{Measuring cosine similarities}\label{ss:cossim}
Through preparation steps described above, we now have a vector representation for each phone: $ D = \{(p_i, \mathbf{r}_i)\}$.
To quantitatively measure whether each quadruplet $\mathbf{p}=(p_1, p_2, p_3, p_4)$ is consistent with the underlying representations, we calculate the average cosine similarity:
\begin{align}
    \overline{\cos}(\mathbf{p}) = \mathbb{E}[\cos(\mathbf{r}_{p_1}, \mathbf{r}_{p_2} + \mathbf{r}_{p_3} - \mathbf{r}_{p_4})],\label{eq:avg_synth}
\end{align}
where the average is calculated by randomly sampling each phone representation from $D$.

To quantify the uncertainty of similarity estimate, we use bootstrapping, similar to \citet{choi2024self}.
We randomly sample $1000$ phones each with replacement, and compute the averaged cosine similarity.
In our preliminary experiments, we found that each bootstrap estimate is stable, resulting in very small across-replicate variance.
As such, we construct the 99\% confidence interval (CI) using 10 replicates, to avoid excessive computation.

\subsubsection{Comparing cosine similarities}\label{ss:cosbaseline}
We compare the above cosine similarities with two baselines.
The \textbf{same-phone baseline} provides an upper bound by measuring the similarity between the same phone $p_1$ from different utterances:
\begin{align}
    \overline{\cos}^+(\mathbf{p}) = \mathbb{E}[\cos(\mathbf{r}_{p_1}, \mathbf{r}'_{p_1})],\label{eq:avg_same}
\end{align}
where $\mathbf{r}'_{p_1}$ is another random sample of $p_1$ from $D$.

Similarly, \textbf{different-phone baseline} measures between $p_1$ and random phone that is not $p_1$:
\begin{align}
    \overline{\cos}^-(\mathbf{p}) = \mathbb{E}[\cos(\mathbf{r}_{p_1}, \mathbf{r}_{\text{not } p_1})].\label{eq:avg_diff}
\end{align}

For a well-structured representation, we expect the similarities to satisfy the ordering:
\begin{align}
    \overline{\cos}^-(\mathbf{p}) < \overline{\cos}(\mathbf{p}) < \overline{\cos}^+(\mathbf{p}).\label{eq:success_ineq}
\end{align}
We assess whether this ordering holds by comparing the confidence intervals (CIs) of the estimates, verifying that the upper CI of the left term is below the lower CI of the right term.
\Cref{eq:success_ineq} is analogous to an ABX test \citep{chaabouni2017learning}, where the target representation is expected to be more similar to another instance of the same phone (\cref{eq:avg_same}) than to its approximation (\cref{eq:avg_synth}), and more similar to its approximation (\cref{eq:avg_synth}) than to representations of other phones (\cref{eq:avg_diff}).

To summarize the behavior over all quadruplets, we define the \textbf{success rate}, which is the proportion of quadruplets whose similarity scores satisfy the ordering in \cref{eq:success_ineq} such that phonological analogies hold:
\begin{align}
    S(Q) = \frac{1}{|Q|}\sum_{\mathbf{p} \in Q} \mathbbm{1}[\overline\cos^-(\mathbf{p})<\overline\cos(\mathbf{p})<\overline\cos^+(\mathbf{p})],
\end{align}
where $\mathbbm{1}$ denotes the indicator function, returning $1$ if the condition holds, and $0$ otherwise.
We additionally evaluate performance using an alternative metric, described in \cref{ss:fragile}.

\subsubsection{Speech representations}\label{ss:models}\label{ss:pooling}
We use spectral representations as baselines.
We employ log mel spectrograms (\textbf{MelSpec}) and mel-frequency cepstral coefficients (\textbf{MFCC}), using the default parameters of librosa \citep{mcfee2015librosa}.

We compare these baselines to three monolingual S3Ms trained on English: \textbf{wav2vec 2.0} \citep{baevski2020wav2vec}, \textbf{HuBERT} \citep{hsu2021hubert}, and \textbf{WavLM} \citep{chen2022wavlm}, using the \textsc{Large} configuration.
We extract 25 layerwise representations per model.
We defer additional details to \cref{ss:s3m-detail}.

Given the speech utterance $\mathbf{x} \in \mathbb{R}^T$ of length $T$, speech representation is a matrix $\mathbf{R}=f(\mathbf{x}) \in \mathbb{R}^{T'\times F}$, where the temporal dimension (number of frames) is reduced to $T' \simeq T / s$ with the model's stride size $s$.\footnote{The evaluated S3Ms use $s=320$ for 16kHz input audio.}
Our goal is to extract a single vector $\mathbf{r} \in \mathbb{R}^F$ for each phone segment, given its start and end times $(t_s, t_e)$.
Following \citet{pasad2021layer,pasad2023comparative}, we conduct average pooling after applying feature slicing.
\cref{ss:feat.vs.audio} conducts additional comparisons on different slicing methods.


\subsection{Results}\label{ss:ex1-result}
\begin{figure}[t]
\centering
  \includegraphics[width=\linewidth]{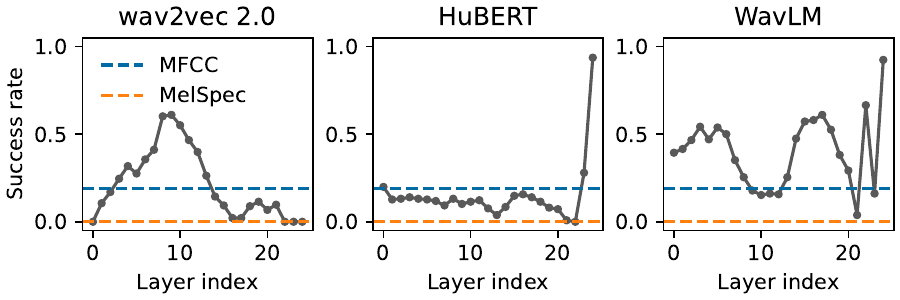}
  \includegraphics[width=\linewidth]{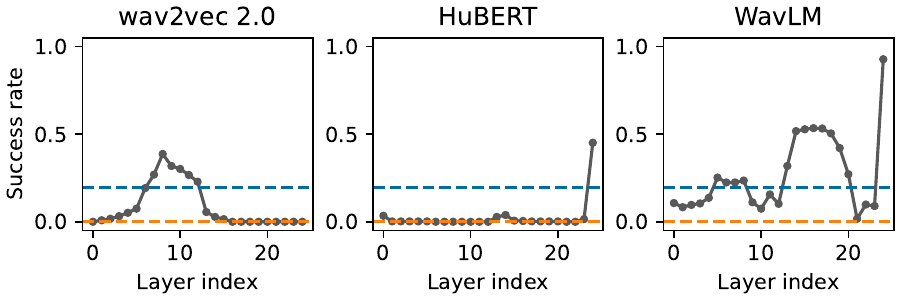}
  \caption{
    Comparing S3Ms with spectral representations on TIMIT (top) and VoxAngeles (bottom).
  }
  \label{fig:model-compare}
\end{figure}

\subsubsection{S3Ms vs. Spectral representations}\label{sss:dir-s3ms}
\Cref{fig:model-compare} compares the success rates of S3Ms and spectral representations on phonological analogies using TIMIT.
The last layer of HuBERT (94\%) and WavLM (92\%), as well as the middle layer of wav2vec 2.0 (61\%) substantially outperforms spectral features MFCC (19\%) and MelSpec (0\%).

Each S3M exhibits distinct layerwise behavior:
wav2vec 2.0 peaks in the middle layer, whereas HuBERT and WavLM peak in the last layer.
This behavior is consistent with prior observations that measure layerwise phonetic information through probing \citep{pasad2023comparative,choi2025leveraging}.
Our results extend these findings by showing that S3Ms exhibit phonological compositionality.

Further, we observe that a greater number of analogies hold in the middle or final layers compared to layer index 0.
We hypothesize that the need for deeper layers suggests that S3Ms benefit from increased contextualization when forming abstract phonological vectors.
We explore this hypothesis further in \cref{sss:dir-consvowel} and \cref{ss:feat.vs.audio}.

Additionally, we compare phone recognizers that are fine-tuned from S3Ms to assess the impact of the phone recognition task on the phonological analogies (\cref{ss:prft}).
We also find that alternative evaluation metrics can yield different layerwise trends in \Cref{fig:model-compare} (\cref{ss:ocs}).

\subsubsection{Do S3Ms generalize to unseen phones?}\label{sss:dir-unseen}
We examine whether phonological analogies hold for phones from unseen languages using VoxAngeles dataset (\Cref{fig:model-compare}).
Of the 468 analogies, 316 (68\%) contain at least one phone that does not exist in the English (TIMIT) phone set.

Consistent with the trends observed in \Cref{sss:dir-s3ms}, S3Ms continue to achieve higher success rates than spectral representations.
In particular, WavLM, HuBERT, and wav2vec 2.0 achieve success rates of 93\%, 45\%, and 39\%, respectively, compared to 19\% for MFCC and 0\% for MelSpec.
This indicates that English-only S3Ms capture phonological structure beyond English-specific phones.

\subsubsection{Vowels vs. Consonants}\label{sss:dir-consvowel}
\begin{figure}[t]
\centering
  \includegraphics[width=\linewidth]{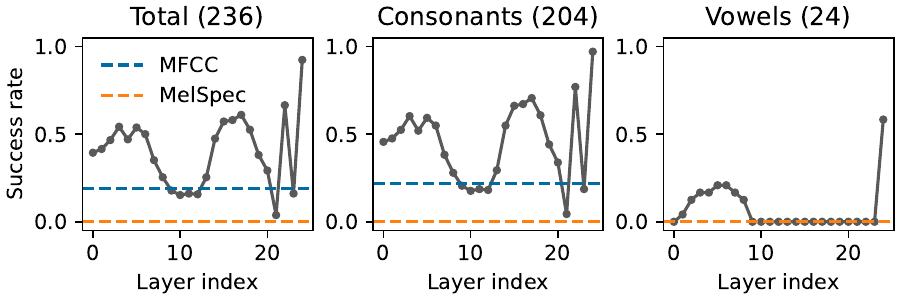}
  \includegraphics[width=\linewidth]{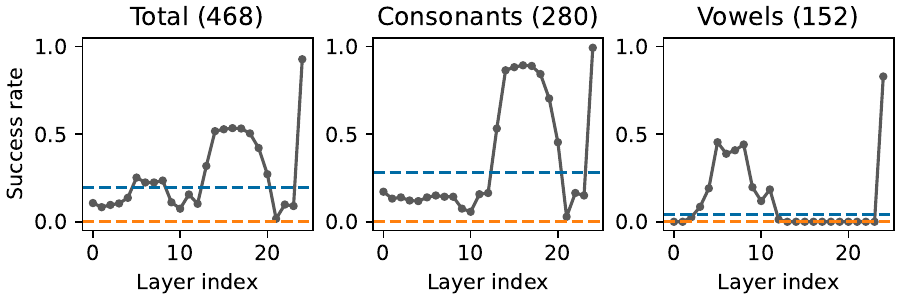}
  \caption{
    Comparing consonant-only and vowel-only quadruplets on TIMIT (top) and VoxAngeles (bottom) for WavLM.
    Number within the parenthesis denotes the number of quadruplets.
    We exclude cases where a quadruplet contains both consonants and vowels.
  }
  \label{fig:cons-vs-vowels}
\end{figure}

We analyze WavLM representations, which achieve the highest overall success rates (\Cref{fig:model-compare}).
We observe three prominent peaks in success rate: (1) a first intermediate peak between layers 0--10, (2) a second intermediate peak between layers 10--20, and (3) the highest peak at the final layer.

To further investigate the origin of different peaks, we separate phonological analogies into vowels and consonants.
As shown in \Cref{fig:cons-vs-vowels}, vowels are associated with the first intermediate peak across both datasets.
In contrast, consonants exhibit more complex behavior: in TIMIT, consonants contribute to both intermediate peaks, whereas in VoxAngeles they are associated with the second intermediate peak.
Nevertheless, both vowels and consonants peak on the final layer.

We suggest that differences may stem from the distinct acoustic-temporal properties of vowels and consonants.
Vowel cues tend to be temporally localized, whereas consonantal cues are often distributed across surrounding segments.\footnote{For example, aspirated plosive [p$^{\text{h}}$] in the word \textit{apply} can be inferred from multiple cues spanning a broader temporal context, including formant transitions in the preceding vowel, burst energy at release, subsequent aperiodic noise, and partial devoicing of the following [l].
}
Given that deeper layers are more likely to leverage broader contextual information, we speculate that phonological features requiring less temporal context often peak earlier in the network (\textit{e.g.}, vowels), and vice versa (\textit{e.g.}, consonants).

In summary, these findings indicate that the strong performance of S3Ms is closely tied to their ability to leverage contextual information.
This conclusion is further supported by experiments that limit the temporal window size (\cref{ss:feat.vs.audio}).
Moreover, results suggest that different temporal complexities are preferentially contextualized at different layers, while the final layer gathers them into a unified representation.

We additionally tested other factors beyond the consonant-vowel distinction that may dictate such layerwise trends.
However, we found that neither individual phonological features (\cref{ss:layerwise-panphon}) nor phonological distances between analogies (\cref{ss:layerwise-pfer}) substantially influence these layer-wise patterns, leaving more underlying causes for future investigation.





\section{Experiment 2: Scale of Phonological Vectors}\label{sec:degree}
In this section, we test the second hypothesis: whether the scale $\lambda$ of the phonological vectors (\cref{eq:weighted-voicing}) correlates with acoustic measurements associated phonological features, by training a vocoder to invert S3Ms.


\begin{table*}[t]
\centering
\resizebox{0.9\textwidth}{!}{%
\begin{tabular}{lcccccccc}
\toprule
\textbf{Phonological feature} 
& High & Low & Back & Round & Nasal & Sonorant & Strident & Voice \\
\midrule
\textbf{Acoustic measurement} 
& F1 & F1 & F2 & F2 & F1BW & HNR & COG & COG \\
\textbf{Expected correlation sign} 
& -- & + & -- & -- & -- & + & + & -- \\
\bottomrule
\end{tabular}
}%
\caption{
Summary of phonological features, their associated acoustic measurements, and expected correlation signs.
The phonologically expected correlation sign is denoted as + (positive) or -- (negative).
Five types of acoustic measurements are used: first formant (F1), second formant (F2), first-formant bandwidth (F1BW), harmonic-to-noise ratio (HNR), and center of gravity (COG).
Further details are provided in \cref{ss:measure}.
}
\label{tb:feat-summary}
\end{table*}

\begin{figure*}[t]
\centering
  \includegraphics[width=\linewidth]{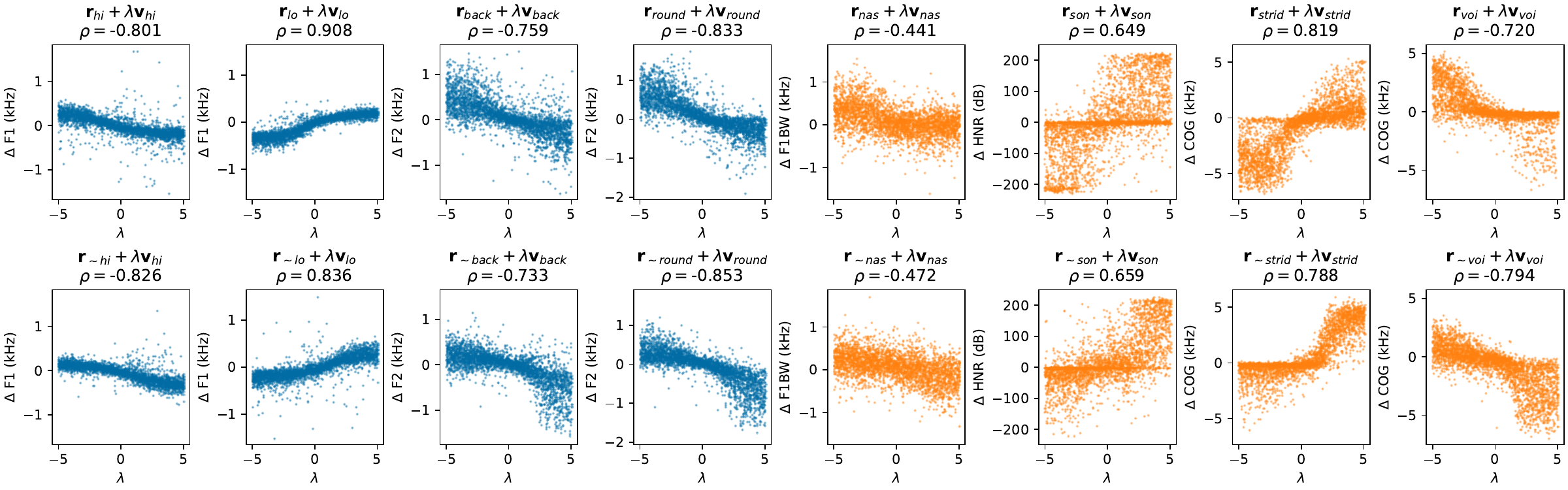}
  \caption{
    Comparison between the phonological vector scale $\lambda$ and the acoustic measurements (\cref{ss:measure}) on TIMIT.
    $\rho$ denotes Spearman's rank correlation coefficient.
    Blue and orange plots indicate vowels and consonants, respectively.
    The empirically observed correlation signs match the theoretical expectations shown in \Cref{tb:feat-summary}.
    Further, plots demonstrate the linearity of phonological vectors, resulting in monotonic (but not necessarily linear) changes in acoustic measurements.
  }
  \label{fig:scatter-timit}
\end{figure*}

\subsection{Method}\label{ss:exp2-method}

\subsubsection{Modifying representations through phonological vectors} \label{ss:modify}
We define each phonological vector $\mathbf{v}$ using the PanPhon features $\mathbf{h}$ in \cref{ss:quads}.
We define the phonological vector as the difference between the mean representations of phones with and without the feature $i$ (not necessarily minimal pairs):
\begin{align}
    \mathbf{v}_i = \mathbb{E}_{\mathbf{h}[i] = +1}[\mathbf{r}] - \mathbb{E}_{\mathbf{h}[i] = -1}[\mathbf{r}].\label{eq:phonovec}
\end{align}
Motivated by the analysis of \cref{sss:dir-consvowel}, we separately compute consonants and vowels, using consonant-derived vectors for consonants, and vice versa.
For example, the voicing vector is obtained by subtracting the averaged representations $\mathbf{r}$ of all voiced consonants ($\mathbf{h}[\text{voi.}]=+1$) with all unvoiced consonants ($\mathbf{h}[\text{voi.}]=-1$).
We additionally analyze the sample efficiency of \cref{eq:phonovec} (\cref{ss:phono-compare}) and compare with single phone pair constructions (\cref{ss:phono-singlepair}).

As mentioned in \cref{eq:weighted-voicing}, we apply vector $\mathbf{v}$ to the frames corresponding to the target phone $p$.
Given its start and end frame indices $t_s', t_e'$ (\cref{ss:pooling}), the modified representation $\tilde{\mathbf{R}}$ is defined as:
\begin{align}
    \tilde{\mathbf{R}}_t = \begin{cases}
        \mathbf{R}_t + \lambda \mathbf{v}  & (t_s' \leq t < t_e')\\
        \mathbf{R}_t & (\text{otherwise}),
    \end{cases}
\end{align}
where the scaling factor $\lambda \in \mathbb{R}$ controls the strength of the modification and $\mathbf{R}_t$ denotes the representation at timestep $t$.
Finally, using the vocoder model $f^{-1}$, we reclaim the expected speech through resynthesis: $\tilde{\mathbf{x}} = f^{-1}(\tilde{\mathbf{R}})$. 
We use the final layer representations from WavLM, as it has been shown to be effective for both TIMIT and VoxAngeles (\Cref{ss:ex1-result}).

\begin{figure*}[t]
\centering
  \includegraphics[width=\linewidth]{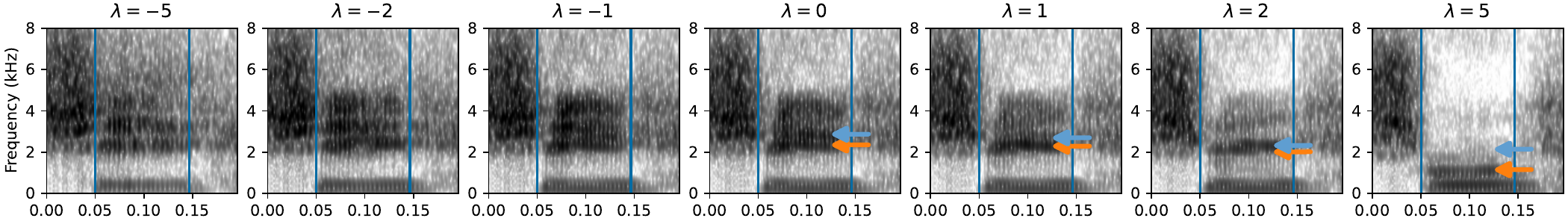}
  \vspace{-2em}
  \caption{
    Applying round vector to front vowel [i], where there is no front rounded vowel in English.
    Orange and blue arrows denote F2 and F3, respectively, which are all decreasing for $\lambda >0$.
  }
  \label{fig:synth-round}
  \vspace{0.5em}

  \includegraphics[width=\linewidth]{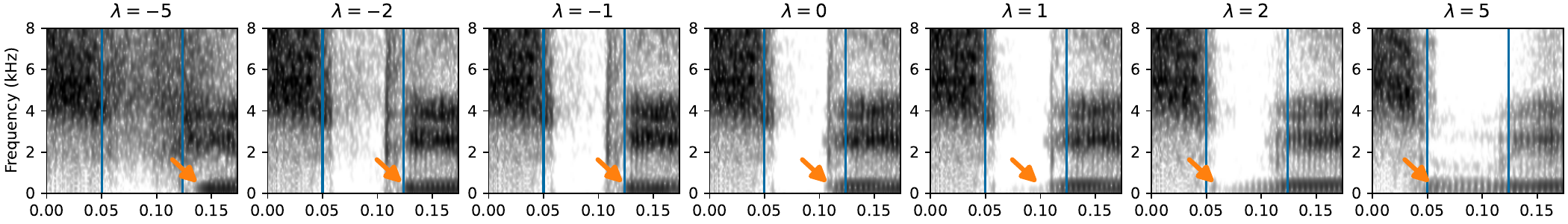}
  \vspace{-2em}
  \caption{
    Applying the voicing vector to phone [b].
    Orange arrows denote the onset of voicing in the subsequent vowel.
    For increasing values of $\lambda$, the voicing onset time is moved earlier, extending into the closure segment for large values of $\lambda$.
  }
  \label{fig:synth-voicing}
  \vspace{0.5em}

  \includegraphics[width=\linewidth]{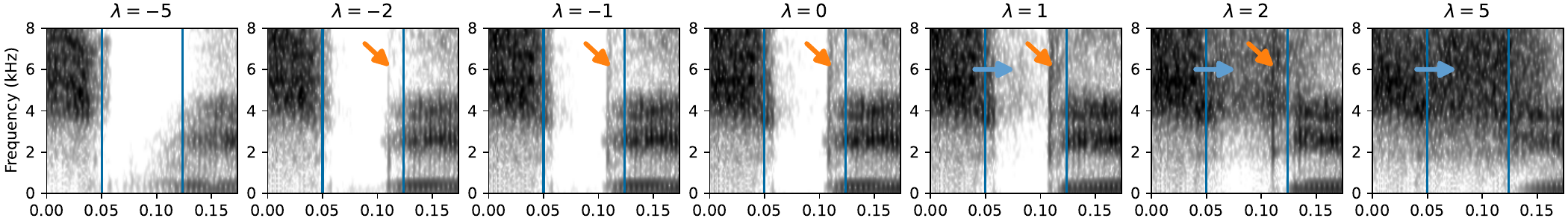}
  \vspace{-2em}
  \caption{
    Applying the strident vector to phone [b].
    Orange arrows point to the burst, where increasing $\lambda$ removes the burst.
    Blue arrows denotes the increasing energy around 4$\sim$8kHz.
  }
  \label{fig:synth-strident}
  \vspace{0.5em}

  \includegraphics[width=\linewidth]{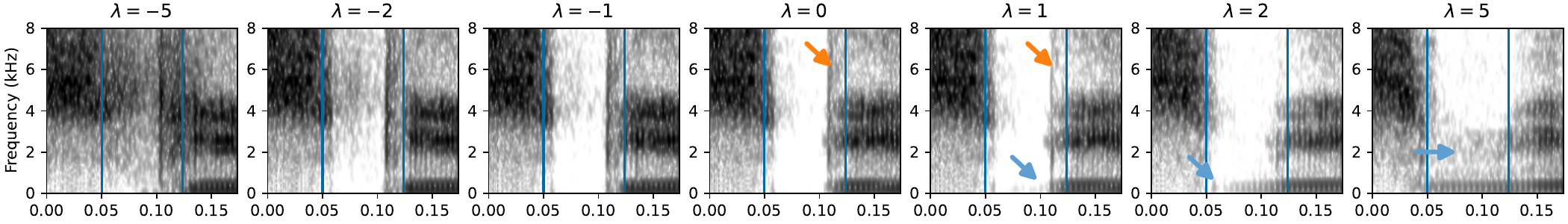}
  \vspace{-2em}
  \caption{
    Applying the nasal vector to phone [b].
    Orange arrows point to the burst, where increasing $\lambda$ removes the burst.
    Blue arrows point to the low frequency murmur introduced with increasing nasalization.
  }
  \label{fig:synth-nasal}
\end{figure*}

\subsubsection{Analyzing modified representations through acoustic measurements}\label{ss:synth-corr-measure}
To assess whether the scaling factor $\lambda$ reflects the degree of associated phonological feature realization, we extract acoustic measurements from the modified representation $\tilde{\mathbf{R}}_t$.
Specifically, we use the vocoder $f^{-1}$ to resynthesize speech $\tilde{\mathbf{x}}$ from $\tilde{\mathbf{R}}$, and then compute the acoustic measurements for the targeted phonological feature.
These measurements quantify how the degree of feature realization in the resynthesized speech varies with respect to $\lambda$.
For example, increasing $\lambda$ for the voicing vector is expected to yield a higher degree of voicing in the resynthesized target phone.

To quantify these effects, we compare acoustic measurements before and after modifying representations with scaled phonological vectors.
We assess the relationship between the scale $\lambda$ and the change of acoustic measurements $\Delta$ using Spearman’s rank correlation coefficient ($\rho$).

\subsubsection{Vocoder training}\label{ss:vocos}
We train two vocoder models based on the Vocos vocoder \citep{siuzdak2024vocos}: an English model using LibriTTS \citep{zen2019libritts}, and a multilingual model using FLEURS-R \citep{ma2024fleurs}.
Vocos mitigates synthesis artifacts commonly introduced by aggressive upsampling and demonstrates robustness to out-of-distribution inputs \citep{siuzdak2024vocos}, making it well suited for analyzing S3M representations.
Details of vocoder training can be found in \cref{ss:vocos-detail}.
We quantitatively evaluate resynthesis quality using acoustic measurements in \cref{ss:vocos-stability}.

\subsubsection{Phonological vectors}\label{sss:synth-phonovectors}
We test eight phonological vectors with well-established corresponding acoustic measurements: \textit{high}, \textit{low}, \textit{back}, and \textit{round} for vowels, and \textit{nasal}, \textit{sonorant}, \textit{strident}, and \textit{voice} for consonants.
We summarize the acoustic measurements on \Cref{ss:measure} and the expected correlation sign in \Cref{tb:feat-summary}.
We additionally analyze the non-orthogonal relationships between these phonological vectors in \cref{ss:phono-compare}.
We also examine the influence of the \textit{round} phonological vector across multiple acoustic measurements in \cref{ss:synth-rounding}.

To estimate the phonological vector $\mathbf{v}$, we use TIMIT train split and a fixed randomly selected subset of languages for VoxAngeles.
For each vector, we modify and resynthesize 3000 utterances drawn from the remaining data, \textit{i.e.}, the TIMIT test split and the remaining VoxAngeles languages.
For each utterance, we sample $\lambda \in U(-5, 5)$ from the uniform distribution and a phone segment (with replacement), modify its representation using $\lambda \mathbf{v}$, and resynthesize using the vocoder $f^{-1}$.

\subsection{Results}
\subsubsection{Quantitative analyses}\label{sss:scatter-analysis}
\Cref{fig:scatter-timit} compares the acoustic measurements of the original and the modified-resynthesized speech on TIMIT.
Across all features, the correlation signs observed in our experiments exactly matches the theoretically expected signs (\Cref{tb:feat-summary}).
Further, we observe consistent and monotonic relationships between the vector scale $\lambda$ and the resulting acoustic measurements.
Our results confirm that the phonological vectors behave in accordance with their intended interpretation as linear directions in the learned representation space, inducing monotonic but not necessarily linear changes in acoustic measurements.
\Cref{fig:scatter-voxangeles} in \cref{ss:voxangeles-synth} further shows that the same trends hold for VoxAngeles, demonstrating the generalization abilities to unseen phones during training.

Further, we found the effects of the phonological vectors are continuous rather than binary.
For example, increasing the scale of the voicing vector does not abruptly toggle voicing on or off.
Instead, it produces smooth shifts in COG, reflecting a graded change in the degree of voicing.
This suggests that S3M representations encode phonological features not merely as categorical contrasts, but as continuous directions.
This property enables fine-grained control over acoustic variation along individual phonological dimensions.

We also observe robust extrapolation well beyond the interpolation range $|\lambda| \leq 1$.
$|\lambda| > 1$ still yields acoustically interpretable outputs, further supporting the linear structure of phonological vectors within the representation space.

There were three exceptions for extrapolation.
For sonorance, the curve is effectively saturated for already-sonorant segments, reflecting the fact that sonorants cannot easily be made ``more sonorant.''
Similarly, voiced consonants and non-stridents also show comparable saturation effects.

\subsubsection{Qualitative analyses}\label{sss:synth-qualitative}
We complement the quantitative results with a qualitative inspection of spectrograms.
For each phonological feature, we resynthesize audio using $\lambda = (-5, -2,-1, 0, 1, 2, 5)$.
We visualize the modified phone with 500ms of context on both sides to show both the local effect and potential coarticulation (\Cref{fig:synth-round,fig:synth-voicing,fig:synth-strident,fig:synth-nasal}).

The \textbf{rounding} vector is applied to the high front unrounded vowel [i] in \Cref{fig:synth-round}.
For $\lambda >0$, all formants are being lowered together, consistent with an acoustic hallmark of lip rounding.
Note that English has no front rounded vowels, demonstrating that rounding vector generalizes to unseen phones.

The \textbf{voicing} vector is applied to the voiced bilabial plosive [b] in \Cref{fig:synth-voicing}.
For $\lambda < 0$, the onset time of voicing after the plosive is increased.
For $\lambda > 0$, the model decreases the voice onset time (VOT).
For large values, the voicing of the subsequent vowel is extended into the closure of the plosive, exhibiting negative VOT.

The \textbf{strident} vector is applied to [b] in \Cref{fig:synth-strident}.
For $\lambda > 0$, frication above 4kHz increasingly emerges, matching the spectral signature of strident fricatives.
Notably, it also removes the burst characteristics of plosives when increasing stridency.
This indicates that S3M representations encode not only static spectral envelopes, but also internal temporal structure (burst vs. frication), and modify them coherently along the phonological dimension.

The \textbf{nasal} vector is applied to [b] in \Cref{fig:synth-nasal}.
Increasing $\lambda$ introduces nasal acoustic cues: weakening of the burst, and introduction of a low-frequency murmur.
As with stridents, these modifications affect both temporal and spectral structure associated with the underlying manner of articulation, not merely its coarse spectral profile.

Taken together, our analyses show that phonological vectors induce phonologically coherent changes in speech representations.
These effects follow expected trajectories for individual phonological features, vary continuously rather than categorically, and in some cases extrapolate in linguistically interpretable ways.
Moreover, the vectors modulate not only spectral envelopes but also temporal cues, indicating that S3M representations encode rich internal structure.
These findings further connect S3M representations to scalar and multi-valued features in phonological theory (see \citet{gnanadesikan1997phonology} for a survey).

\section{Related works}
\textbf{Word analogies.}
\citet{Mikolov2013LinguisticRI} demonstrated linear analogies in word embeddings, showing syntactic and semantic relations through vector arithmetic.
\citet{kim2013deriving} extended to continuous semantic scales for adjectives, which is analogous to scaling phonological vectors with $\lambda$.
Also, \citet{pennington2014glove} use analogical tasks to compare embeddings, and \citet{ethayarajh2019towards} provide a mathematical explanation for linear analogies, motivating our work for phonological analogies.
Additionally, \citet{levy2014linguistic} raised concerns about word analogy evaluation, which motivated \citet{fournier2020analogies} to propose alternative methods based on comparing relational directions rather than representations directly.

\textbf{Linear representation hypothesis} (LRH) suggests that human-interpretable features are linearly represented within models' hidden representations \citep{elhage2022toy,park2024linear,modell2025origins}, theoretically supporting the existence of phonological vectors.
LRH is also closely related to steering vectors  \citep{subramani2022extracting,turner2023steering}, which control model behavior by adding interpretable vectors to model representations, motivating our controllable speech synthesis.

\textbf{Speech model interpretability.}
Prior work mainly focused what information is encoded, \textit{i.e.}, spectral \citep{choi2022opening}, phonetic \citep{choi2024understanding}, articulatory \citep{cho2023sslart,cho2024ssluniart}, syllabic \citep{baade2025syllablelm,chosylber}, lexical \citep{peng2022word,pasad2024self}, syntactic \citep{shen2023wave}, and semantic information \citep{pasad2024self}.
More closely related to our work, several studies examined how information is encoded: layerwise analyses revealed linguistic hierarchy \citep{pasad2021layer,pasad2023comparative}, representations store information within similarities \citep{choi2022opening,choi2024self}, and hierarchy among similarities \citep{abdullah2023information,choi2025leveraging}.

\textbf{Phonological analogies.}
For S3Ms, \citet{gauthier2025emergent} showed morphological inflection induces linear geometry and \citet{nakamura2025discrete} identified certain phonological axes within S3Ms.
\citet{li2021hierarchical} also analyzed phone recognition models and identified voicing and aspiration vectors.
\citet{chaabouni2017learning} used phonological analogies and ABX tasks as evaluation tools for multimodal speech representations.
Others also leveraged phonological analogies in text settings: \citet{silfverberg2018sound} showed that phonemic text embeddings can learn phonological vectors without supervision and \citet{zouhar2024pwesuite} used phonological analogies to evaluate them.
In contrast, our work provides a large-scale cross-lingual analysis on S3Ms and further explores the scale of phonological vectors.

\textbf{Controllable speech synthesis} approaches often adopt interpretable features, including phonetic posteriorgrams \citep{zhao2019foreign,morrison2024fine}, articulatory features \citep{cho2024coding,krug2025precisely}, or phonological features \citep{staib2020phonological,taannander2024beyond}, to enable interpretable control.
While they rely on explicitly designed representations, our phonological vectors are driven from self-supervision.
While we do not directly compare synthesis performance, we expect our emergent phonological vectors from S3Ms can be leveraged in future work on such applications.

\section{Conclusion}
We show that S3Ms, trained only on speech without phonological supervision, learn \textit{linearly composable} and \textit{scalable} phonological vectors.
These findings advance both speech processing and linguistics by clarifying the internal structure of S3M representations and refining our understanding of phonological features.
In speech processing, our findings enable intuitive interpretations of S3M representations and fine-grained control of speech synthesis along phonological dimensions.
In linguistics, they provide empirical evidence that phonological features can emerge from acoustic regularities and motivate a view of phonological features as continuous rather than strictly binary.

\section*{Limitations}
This work explored only a limited region of the space of possibilities.
Only a small number of S3Ms were investigated.
Different models behaved differently and it is not possible to isolate what the causes of these differences are on the basis of the results reported here.
Although there can be multiple possible systems of articulatory and acoustic features, our studies only investigated the feature system assumed by PanPhon.
This makes it difficult to draw firm conclusions about whether what is important, in extracting phonological vectors, is identifying a coherent feature that minimally delineates a phonological natural class or simply identifying a consistent phonetic difference.
Finally, synthesis results are influenced not only by the S3Ms but also by the vocoder.
Since we evaluate only a single vocoder, some observed behaviors may reflect vocoder-specific characteristics rather than properties of the S3Ms alone.

\section*{Ethics statement}
Our work focuses on phonological understanding of the self-supervised speech representations through the lens of vector arithmetic.
All of our experiments are conducted using publicly available datasets (\cref{ss:datasets}), which were collected and released under licenses appropriate for research use.
We do not collect new data, nor do we include personally identifiable information beyond what is already present in these datasets.

While our speech synthesis experiments (\cref{sec:degree}) are intended solely for scientific analysis, they may have broader societal implications if misused.
We do not evaluate or claim applicability to the generation of misleading or deceptive content.
We release code, demo, and models for reproducibility and research purposes only, and we encourage future work to consider appropriate safeguards when applying similar techniques in downstream or user-facing systems.

\section*{Use of AI assistants}
AI assistants were used in the preparation of this manuscript.
Specifically, they were employed primarily for code auto-completion, minor text editing, and grammar polishing.
Nevertheless, all scientific content, code implementations, results, and analyses were conceived, verified, and finalized by the authors.

\ifanon
\else
\section*{Acknowledgments}
This material is based upon work supported by the National Science Foundation under Grant No. 2238605.
We are grateful to Kalvin Chang, Christina Bjorndahl, the anonymous ACL reviewers, and the area chair for their helpful comments.
\fi

\bibliography{custom}

\newpage
\appendix

\section{Additional details}
\subsection{Item- and offset-based analogy tests}\label{ss:fragile}
For the word analogy $a:b=a':b'$, \citet{Mikolov2013LinguisticRI} evaluated using item-based analogy tests, \textit{i.e.}, testing whether $b' \simeq a' - a + b$.
Subsequent work has shown that offset-based tests, which assess whether $a-a' \simeq b-b'$, provide a more robust measure of linguistic regularity \citep{levy2014linguistic,fournier2020analogies}.
In this work, we adopt both perspectives: we primarily use item-based tests in \cref{sec:direction}, and employ offset-based tests as a secondary analysis in \cref{ss:ocs}.

The choice of item-based analogy test (success rates) in \cref{sec:direction} is motivated by the characteristics of speech, where the number of paired phones associated with a phonological feature is much smaller compared to the number of words within a word relation.
As a result, offset-based tests on TIMIT and VoxAngeles discard many analogies (36 quadruplets compared to 236 in TIMIT, and 112 compared to 468 in VoxAngeles), which prevents further analyses such as comparisons between vowels and consonants (\cref{sss:dir-consvowel}).
Further, unlike in text, each phone in speech is realized through multiple utterances.
Accordingly, quantities such as $a' - a + b$ do not correspond to a single point estimate but rather to a distribution, for which we report confidence intervals on success rates.

\subsection{Details on S3Ms}\label{ss:s3m-detail}
In \cref{sss:dir-s3ms}, we compared three widely used S3Ms.
We use the \textsc{Large} configuration for all models, consisting of 7 layers of 1D CNNs followed by 24 transformer blocks, a total of approximately 300M parameters.
We extracted representations from each of the 24 transformer layers (index 1$\sim$24) as well as one from the CNNs (index 0), yielding 25 layerwise representations per model.

\textbf{wav2vec 2.0} \citep{baevski2020wav2vec} is trained on 60k hours of English read speech from LibriLight audiobook dataset \citep{kahn2020libri}.
The model is trained using a contrastive loss with on-the-fly learnable codebooks.

\textbf{HuBERT} \citep{hsu2021hubert} is also trained on LibriLight but with a different training strategy.
It uses a predictive loss, where the targets are k-means cluster assignments, either MFCC or the previous training iteration of HuBERT-base.

\textbf{WavLM} \citep{chen2022wavlm} additionally incorporates GigaSpeech \cite{chen21o_interspeech} and VoxPopuli \cite{wang-etal-2021-voxpopuli}, which include both read and spontaneous speech.
Its loss is similar to HuBERT’s, but it includes an additional speech denoising objective.

\subsection{Acoustic measurements}\label{ss:measure}
In \Cref{sec:degree}, we compare eight phonological features and their corresponding acoustic measurements.
We use Parselmouth \citep{jadoul2018introducing}, a Python interface to Praat \citep{boersma2007praat}, to compute the measurements.

\textbf{Formants}, \textit{i.e.}, resonant frequencies of the vocal tract, are used for measuring vowels \citep{ladefoged1996elements,ladefoged2012vowels}.
The first formant (F1) is used for the vowel height (high or low), such that high vowels show lower F1, and vice versa.
Second formant (F2) is used for backness and roundness, such that back vowels and round vowels have lower F2.

\textbf{Bandwidth of F1 (F1BW)}, \textit{i.e.}, the frequency range around F1 with significant energy, is used for measuring nasality \citep{pruthi2007acoustic}.
Nasal sounds exhibit increased damping from nasal cavities, resulting in spreading out the formant energy; a broader F1 bandwidth.

\textbf{Center of gravity (COG)}, \textit{i.e.}, amplitude-weighted average frequency of the spectrum, is used for measuring voicing \citep{bjorndahl2018story} and stridents \citep{strevens1960spectra}.
For voiced sounds, COG decreases due to the presence of the voicing bar, whereas it increases for stridents due to frication in the upper frequency range.


\textbf{Harmonics-to-noise ratio (HNR)}, \textit{i.e.}, the ratio of periodic (harmonic) energy to aperiodic (noise) energy, is used for measuring sonorance \citep{komatsu2002multi}.
Sonorant sounds tend to exhibit higher HNR due to their periodic structure.

\subsection{Details on vocoder training}\label{ss:vocos-detail}
To train the neural vocoder $f^{-1}$, we follow the overall setup of Vocos \citep{siuzdak2024vocos}, with modification to the network architecture and training configuration.
We slightly modify the network architecture dimensions, such that we can use either WavLM or MFCC representations as input.
For faster convergence, we increase both the batch size and the learning rate by a factor of eight relative to the original configuration.

We use two speech synthesis datasets to train a vocoder.
\textbf{LibriTTS} \citep{zen2019libritts} is a multi-speaker English dataset with 585h of read speech.
\textbf{FLEURS-R} \citep{ma2024fleurs} contains 1.3kh of enhanced read speech in 102 languages.

\section{Additional results}
\subsection{Layerwise analysis via offset-based test}\label{ss:ocs}
\subsubsection{Settings}
For the offset-based test, we use the pairing consistency score (PCS) \citep{fournier2020analogies}.
PCS measures the separability between offsets drawn from the same relation and offsets drawn from mismatched relations using a binary classification criterion.

We construct PCS categories by grouping phone pairs that share the same phonological feature differences.
For each category, we compute representation offsets between phone pairs belonging to the same category (correct offsets).
These are contrasted with mismatched offsets, which are formed by shuffling the second phone in each pair.
We use averaged representations per each phone to compute their offsets.

PCS is computed by treating the offset similarity as a binary classification problem, where correct offsets are labeled as positive instances and mismatched offsets as negative instances.
We report the area under the receiver operating characteristic (ROC) curve as the evaluation metric.
A random baseline with no relational consistency yields a PCS of $0.5$.

\begin{figure}[t]
\centering
  \includegraphics[width=\linewidth]{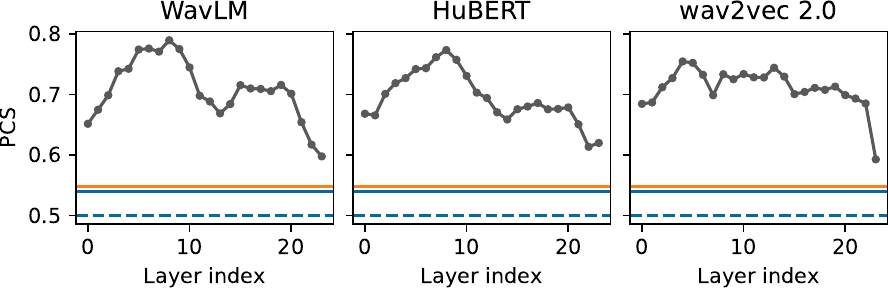}
  \includegraphics[width=\linewidth]{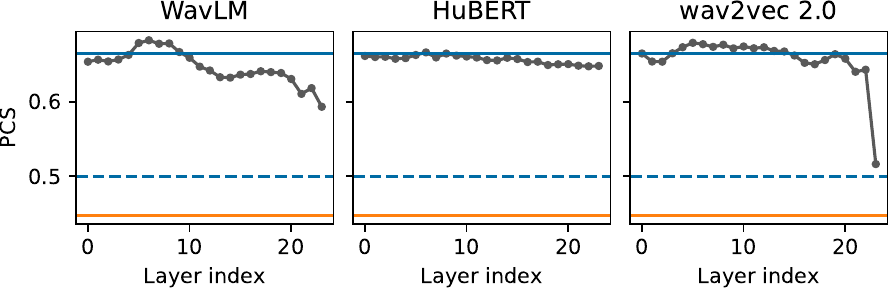}
  \includegraphics[width=\linewidth]{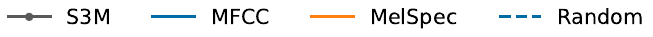}
  \caption{
    Comparing pairing consistency score (PCS) of S3Ms and spectral representations using TIMIT (upper) and VoxAngeles (lower).
  }
  \label{fig:pcs}
\end{figure}

\subsubsection{Results}
\Cref{fig:pcs} presents the layerwise PCS results for S3Ms and spectral representations, and we compare these findings with the success rate results in \cref{sec:direction}.
Consistent with \cref{sss:dir-s3ms}, the PCS results confirm the superiority of S3Ms over spectral representations across layers.
Moreover, PCS scores generally increase in later layers relative to layer 0, suggesting increased contextualization in deeper representations.
In line with \cref{sss:dir-unseen}, we also observe non-negligible performance on unseen phones, although performance remains lower than that achieved on seen phones.

Despite these overall consistencies, the detailed layerwise trends differ from those observed using success rates.
Unlike the patterns reported in \cref{sss:dir-s3ms}, PCS exhibits relatively similar behavior across layers, with peak performance typically occurring in intermediate layers rather than at the final layer.
This suggests that the most informative representations for offset-based relational consistency may reside in middle layers rather than the last layer with the sudden peak of success rates.
We leave a deeper investigation of the discrepancies between different evaluation metrics for future work.

\subsection{Feature vs. Audio slicing}\label{ss:feat.vs.audio}
\subsubsection{Settings}
We compare two common pooling methods for obtaining a phone vector $\mathbf{r}$ from the representation matrix $\mathbf{R} = f(\mathbf{x})$.

\textbf{Feature slicing} \citep{pasad2021layer,pasad2023comparative} performs temporal average pooling on the slice of the representation matrix:
\begin{align}
    \mathbf{r} = \texttt{avgpool}(f(\mathbf{x})[t_s' :t_e']),
\end{align}
where $t_s' = \lfloor t_s/s \rfloor$ and $t_e' = \lceil t_e/s\rceil$ are the corresponding indices after temporal downsampling.
As transformer layers have a global receptive field, it yields a contextualized representation, even though prior work shows that it tends to be locally dominated \citep{baevski2020wav2vec,hsu2021hubert}.

\textbf{Audio slicing} \citep{choi2024understanding,choi2024self}, in contrast, slices the waveform directly:
\begin{align}
    \mathbf{r} = \texttt{avgpool}(f(\mathbf{x}[t_s:t_e])).
\end{align}
This method restricts the temporal receptive field strictly to the phone segment, removing the neighboring context.
Empirically, this has been shown to lead to clearer separation in cosine similarity comparisons \citep{choi2024self}.
If the model requires minimum window size $w > (t_e - t_s)$, we ensure the sliced audio is at least length $w$ by adding equal margins around the segment.

\begin{figure}[t]
\centering
  \includegraphics[width=\linewidth]{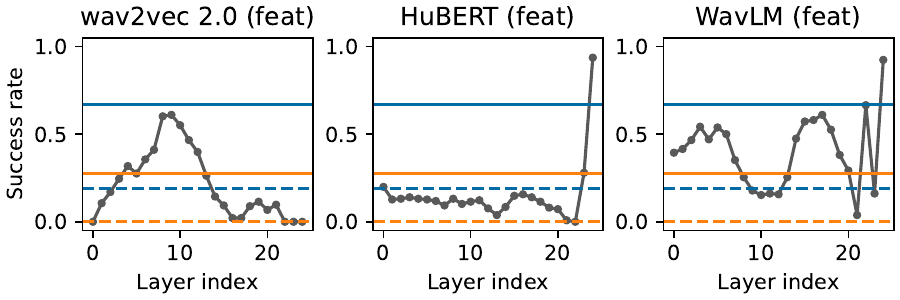}
  \includegraphics[width=\linewidth]{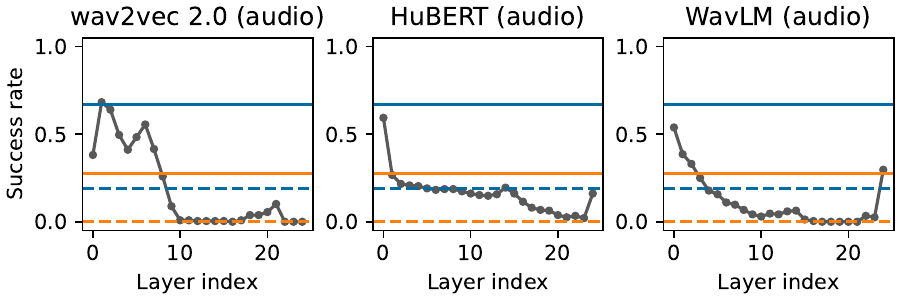}
  \includegraphics[width=\linewidth]{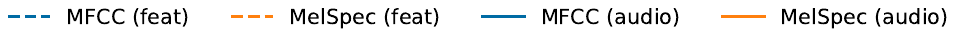}
  \caption{
    Comparing success rates for S3Ms with spectral representations using TIMIT.
    We denote feature and audio slicing as feat and audio, respectively.
  }
  \label{fig:slice}
\end{figure}

\begin{figure}[t]
\centering
  \includegraphics[width=\linewidth]{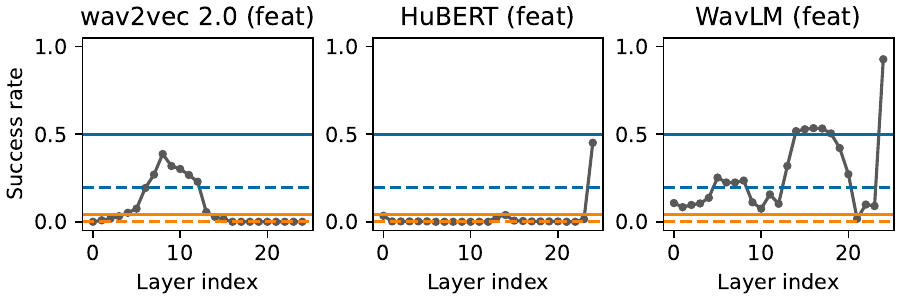}
  \includegraphics[width=\linewidth]{pdfs/model-comparison-legend.pdf}
  \caption{
    Comparing success rates for feature sliced S3Ms with spectral representations using VoxAngeles.
    We denote feature and audio slicing as feat and audio, respectively.
  }
  \label{fig:slice-voxangeles}
\end{figure}

\subsubsection{Results}
In \Cref{fig:slice,fig:slice-voxangeles}, we observe that feature slicing is more effective than audio slicing for S3Ms, whereas the opposite holds for spectral representations.
Overall, feature-sliced WavLM last-layer representations shows strongest performance overall, where we primarily use for all the experiments.

Comparing feature and audio slicing for S3Ms further supports our conclusion in \cref{sss:dir-consvowel}, where audio slicing removes contextualization by limiting the temporal receptive field size.

In contrast, audio slicing is beneficial for spectral representations.
Especially, MFCC reaches a success rate of 67\% and 50\% for TIMIT and VoxAngeles, respectively, which is surprisingly high.
We suspect the difference comes from spectrum magnitude normalization.
For feature slicing, spectral representations' magnitude is normalized per utterance, where audio slicing directly normalizes the magnitude for each phone segment, likely leading to performance improvement.

However, as we show in \cref{sss:dir-cossim}, MFCC representations' cosine similarities tend to collapse toward $1.0$, making them substantially harder to utilize, leading to worse synthesis performance in \cref{ss:mfcc-synth}.

\subsection{S3Ms avoid anisotropic collapse} \label{sss:dir-cossim}
\subsubsection{Settings}
To observe the absolute similarity values, we define the \textbf{averaged similarity}:
\begin{align}
    \overline{C}(A) = \frac{1}{|A|}\sum_{\mathbf{p} \in A} \overline\cos(\mathbf{p}),
\end{align}
and estimate a CI over the set of quadruplet-wise similarities.

\begin{figure}[t]
\centering
  \includegraphics[width=\linewidth]{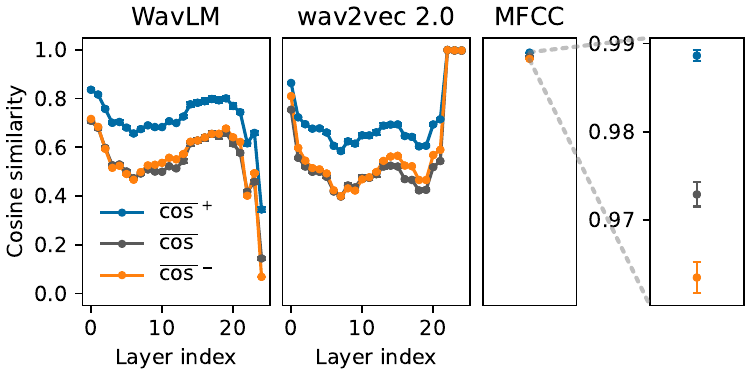}
  \includegraphics[width=\linewidth]{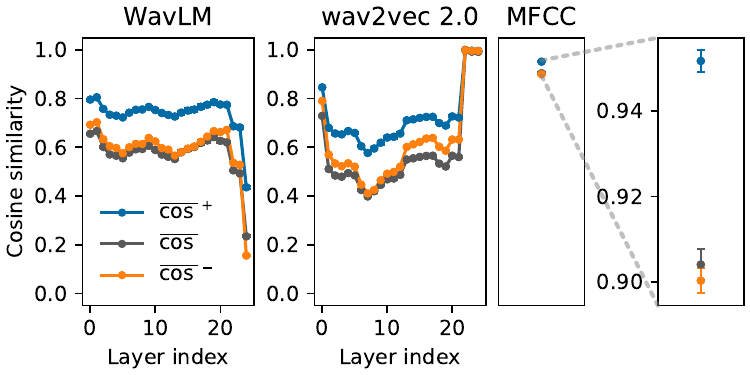}
  \caption{
    Comparing averaged similarities $\overline C$ for feature sliced S3Ms with audio sliced MFCC on TIMIT (upper) and VoxAngeles (lower).
    We calculate the 99\% CI considering quadruplet-wise cosine averages.
  }
  \label{fig:cossim}
\end{figure}

\subsubsection{Results}
\Cref{fig:cossim} shows the averaged cosine similarities for WavLM, wav2vec 2.0, and MFCC.
For later layers of wav2vec 2.0, all similarities approach $1.0$, reflecting anisotropic collapse \cite{ethayarajh2019HowCA}, resulting in reduced success rates in \Cref{fig:model-compare}.
MFCCs also behave similarly, with much smaller margins compared to the last layer of WavLM.
On the other hand, WavLM does not show collapsing behavior, likely leading to more easily usable phonological vectors for synthesis in \cref{sec:degree}.

\subsection{Impact of fine-tuning with phone recognition task}\label{ss:prft}
\subsubsection{Settings}
We compare XLSR-53, a multilingual S3M, and their fine-tuned variants for phone recognition, Wav2vec2Phoneme and MultIPA.
To compare these three models, we use the same settings of \cref{sss:dir-s3ms}.

\textbf{XLSR-53} \citep{conneau2021xlsr} is trained with the wav2vec 2.0 contrastive objective on multilingual datasets spanning 53 languages from CommonVoice \cite{ardila-etal-2020-common}, Multilingual LibriSpeech \cite{pratap20_interspeech}, and BABEL \cite{gales2014speech}, total of 56k hours.

\textbf{Wav2vec2Phoneme} \citep{xu2022simple} fine-tunes XLSR-53 for phone recognition using a CTC loss.
The model is trained on the same datasets, using automatically generated phonemic transcriptions from multilingual G2P systems.

\textbf{MultIPA} \citep{taguchi2023universal} in a similar manner to Wav2vec2Phoneme, but uses a smaller subset consisting of seven languages.

\begin{figure}[t]
\centering
  \includegraphics[width=\linewidth]{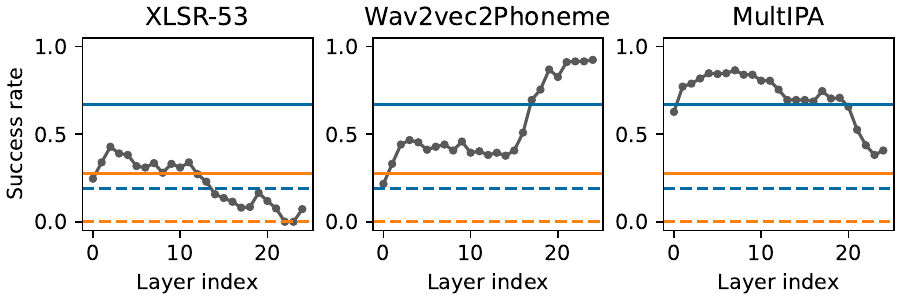}
  \includegraphics[width=\linewidth]{pdfs/model-comparison-legend.pdf}
  \caption{
    Comparing pre-trained S3M (XLSR-53) and fine-tuned phone recognition models (Wav2vec2Phoneme and MultIPA) on TIMIT.
  }
  \label{fig:pr}
\end{figure}

\subsubsection{Results}
\Cref{fig:pr} compares XLSR-53 with their fine-tuned phone recognizer counterparts.
Both fine-tuned models improve success rates relative to XLSR-53 across nearly all layers, indicating that fine-tuning for phone recognition generally strengthens phonological structure.
However, two models differs in their layerwise behavior.
Because the final layer precedes the CTC head, it is optimized to produce representations that linearly separate phones.
Wav2vec2Phoneme shows a sharp increase in success rate in the deeper layers, suggesting that it develops more abstract phonological structure, whereas MultIPA’s improvements are concentrated in earlier layers.
This contrast may be related to differences in language coverage.
As Wav2vec2Phoneme is trained on substantially more languages, it may encourage the model to learn more abstract, general phonological structure, while MultIPA's smaller language set may reduce the pressure for such abstraction.
An interesting avenue for future work is to explore phone recognition as post-training strategies for S3Ms, promoting the emergence of abstract phonological vectors.

\subsection{Different phonological features does not lead to different layerwise trends}\label{ss:layerwise-panphon}

\subsubsection{Settings}
We observe the success rates of individual phonological features on both TIMIT and VoxAngeles.
For each feature, we consider only the quadruplets for which the phonological vector in \cref{eq:analogies} is nonzero, \textit{i.e.}, where either $\mathbf{h}_{p_1}[i] \neq \mathbf{h}_{p_2}[i]$ or $\mathbf{h}_{p_1}[i] \neq \mathbf{h}_{p_3}[i]$ for the feature $i$ under consideration.
Following \cref{sss:dir-consvowel}, we evaluate consonants and vowels separately when computing feature-wise success rates, where quadruplets that contain a mixture of consonants and vowels are excluded from the analysis.

\begin{figure}[t]
  \centering
  \includegraphics[width=\linewidth]{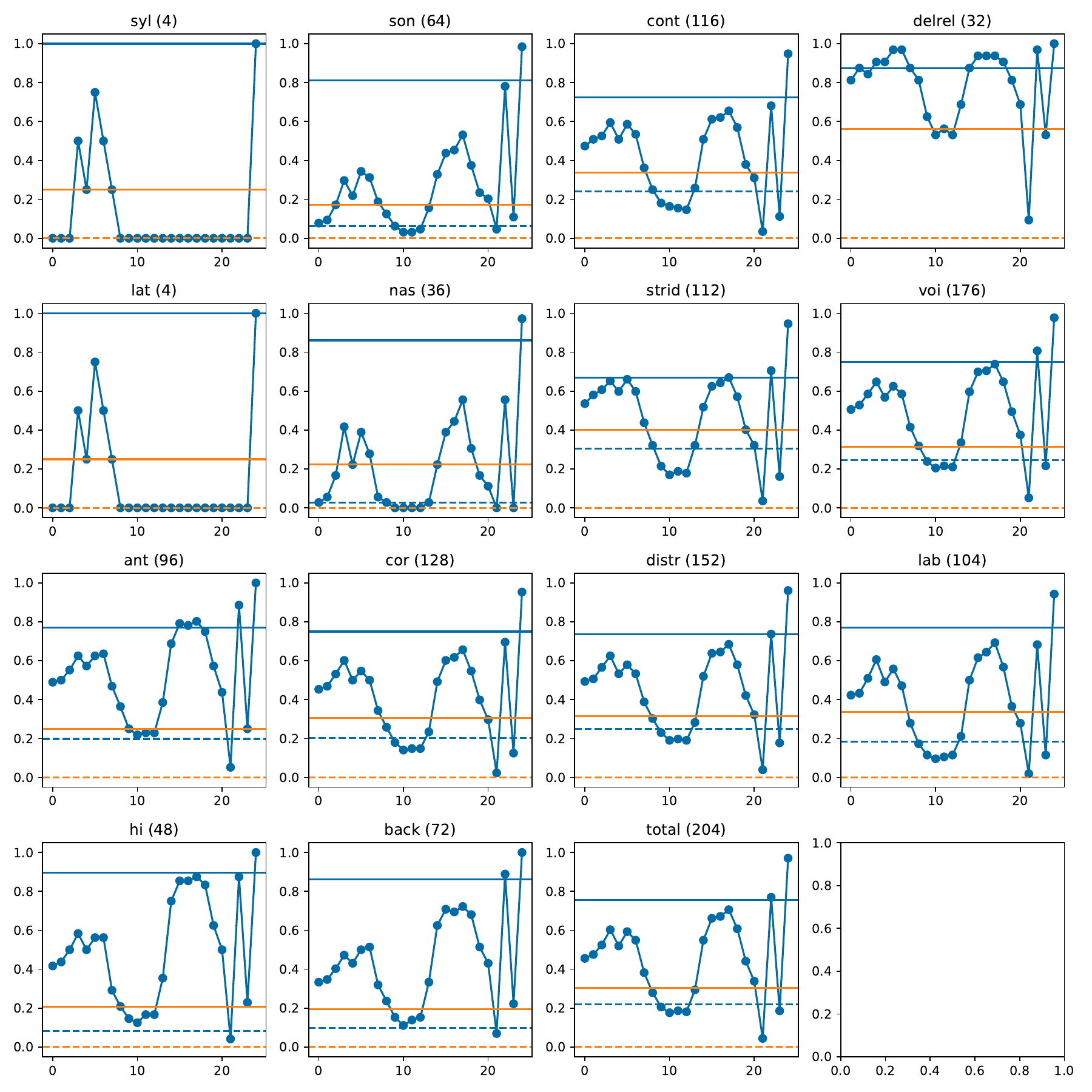}
  \includegraphics[width=\linewidth]{pdfs/model-comparison-legend.pdf}
  \caption{Phonological feature-wise success rates for consonants on TIMIT using WavLM representations.}
  \label{fig:panphon-timit-cons}
  \includegraphics[width=\linewidth]{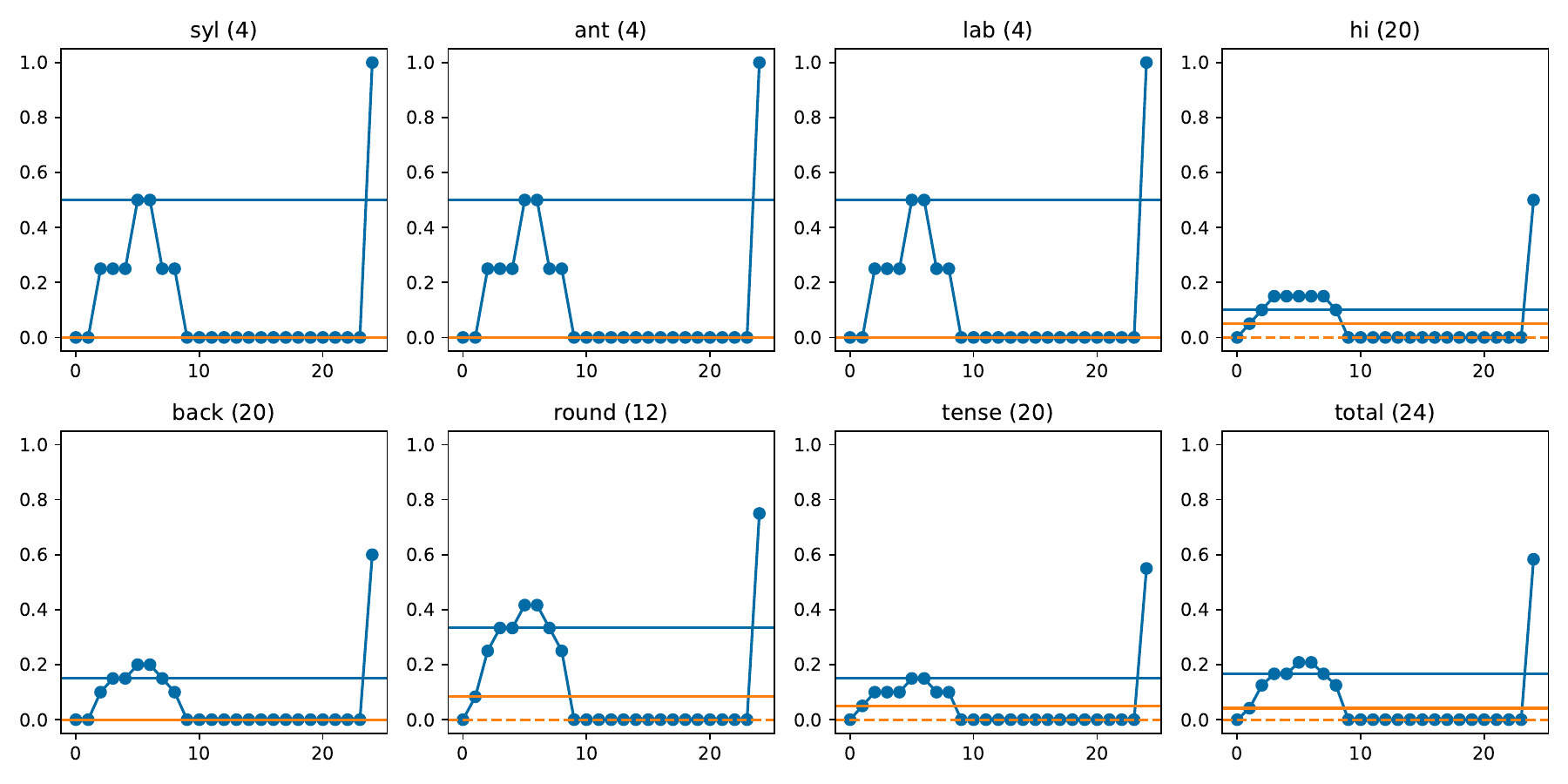}
  \includegraphics[width=\linewidth]{pdfs/model-comparison-legend.pdf}
  \caption{Phonological feature-wise success rates for vowels on TIMIT using WavLM representations.}
  \label{fig:panphon-timit-vowel}
\end{figure}

\begin{figure}[t]
\centering
  \includegraphics[width=\linewidth]{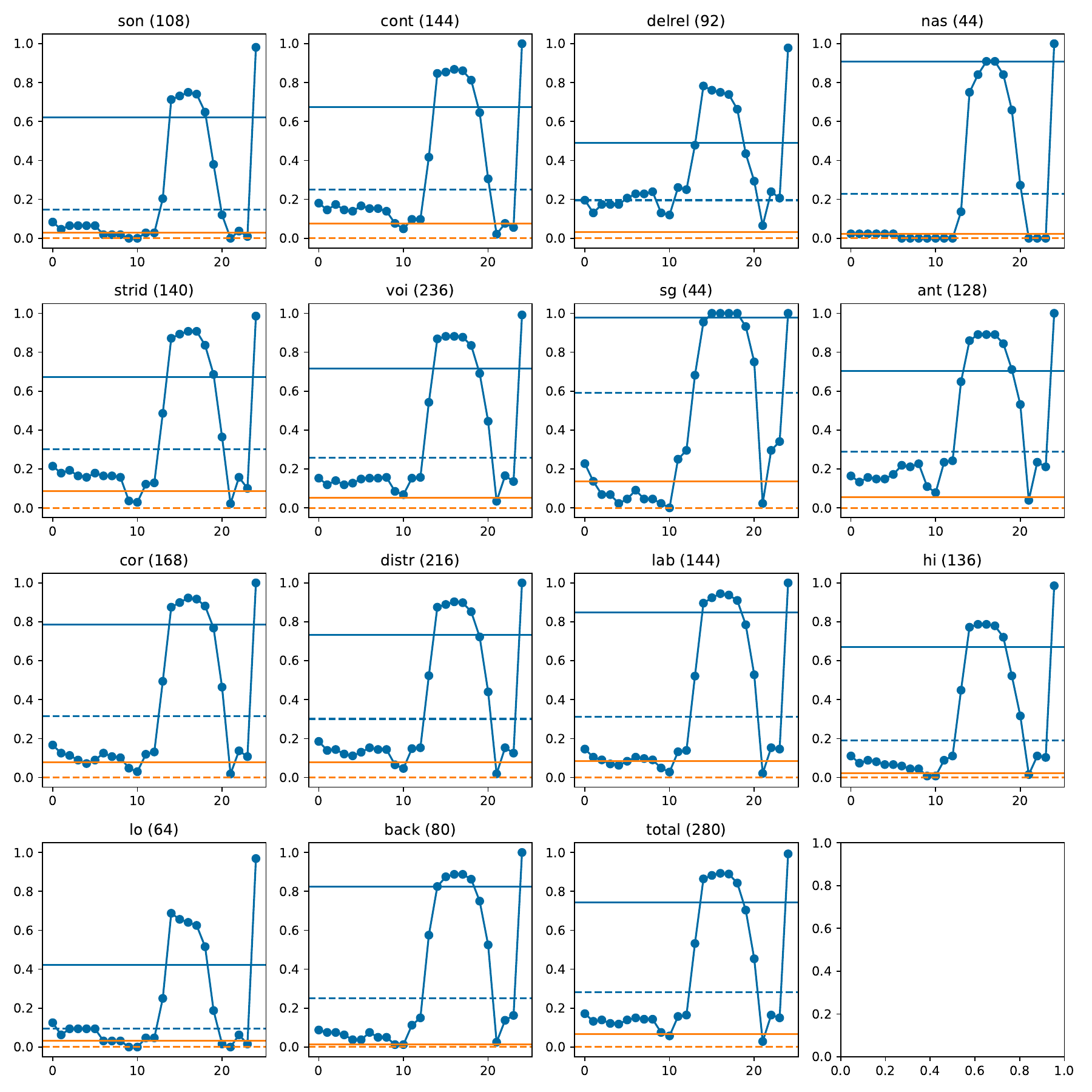}
  \includegraphics[width=\linewidth]{pdfs/model-comparison-legend.pdf}
  \caption{Phonological feature-wise success rates for consonants on VoxAngeles using WavLM representations.}
  \label{fig:panphon-voxangeles-cons}
  \includegraphics[width=\linewidth]{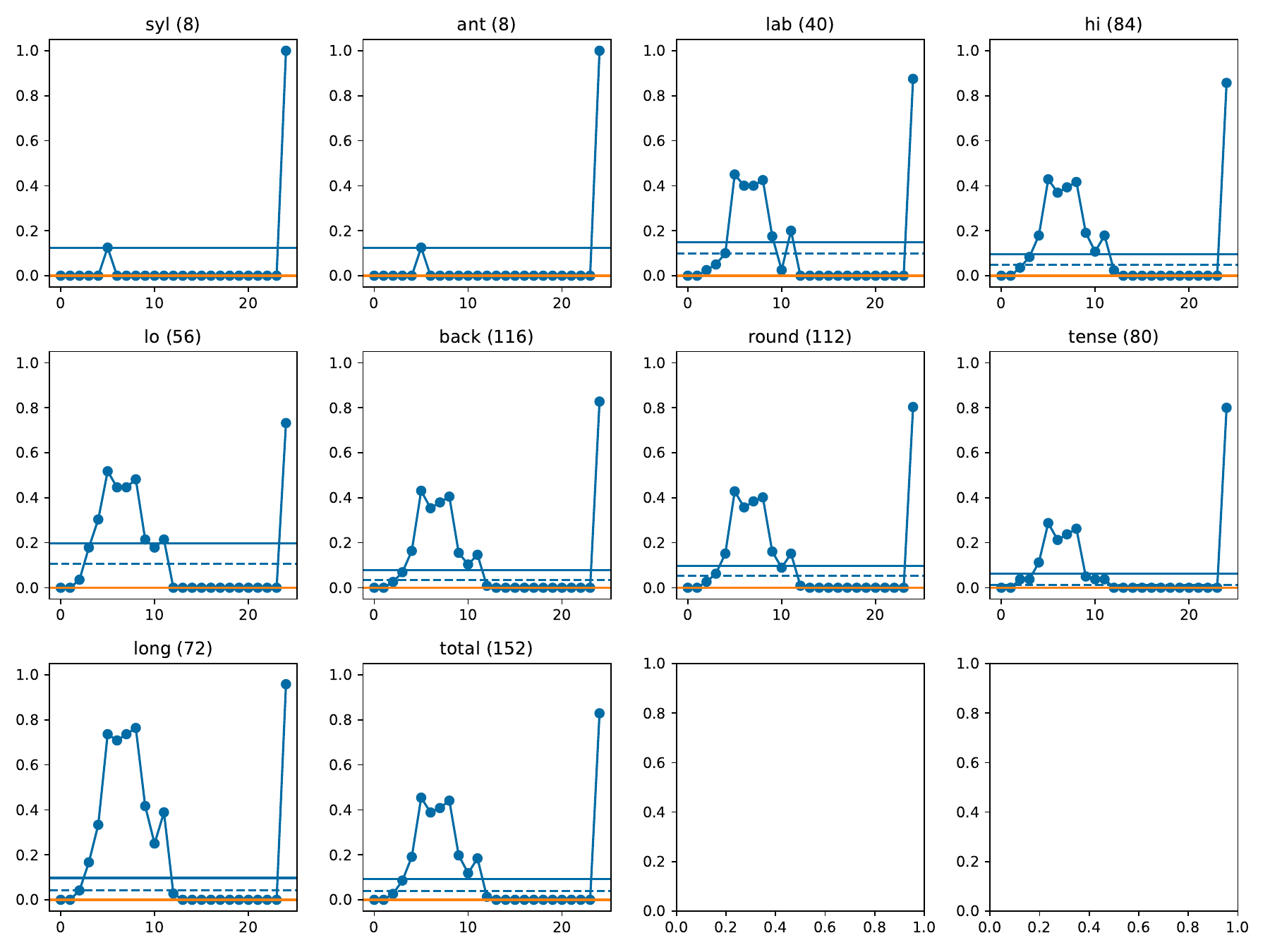}
  \includegraphics[width=\linewidth]{pdfs/model-comparison-legend.pdf}
  \caption{Phonological feature-wise success rates for vowels on VoxAngeles using WavLM representations.}
  \label{fig:panphon-voxangeles-vowel}
\end{figure}

\subsubsection{Results}
For both TIMIT (\Cref{fig:panphon-timit-cons,fig:panphon-timit-vowel}) and VoxAngeles (\Cref{fig:panphon-voxangeles-cons,fig:panphon-voxangeles-vowel}), the layerwise trends are largely consistent across phonological features, with no clear feature-specific differences.
One exception is the syllabic (syl) and lateral (lat) in \Cref{fig:panphon-timit-cons}.
However, the four samples are the permutations of the quadruplet of \textipa{[l]}, \textipa{[n]}, \textipa{[\s{l}]}, and \textipa{[\s{n}]}, where the syllabic consonants behave similar to vowels in terms of their reduced temporal variation.

\subsection{Phonological distances does not lead to different layerwise trends}\label{ss:layerwise-pfer}
\subsubsection{Settings}
We observe the success rates on both TIMIT and VoxAngeles per phonological distance.
Phonological distance between two phones is defined as the number of differing phonological features.
For each quadruplet, we compute the distances between $(p_2, p_3)$ and $(p_2, p_4)$, and assign the quadruplet to a distance bin based on the maximum of these two values.
It ensures that each quadruplet is grouped according to its most phonologically divergent contrast.

\begin{figure}[t]
\centering
  \includegraphics[width=\linewidth]{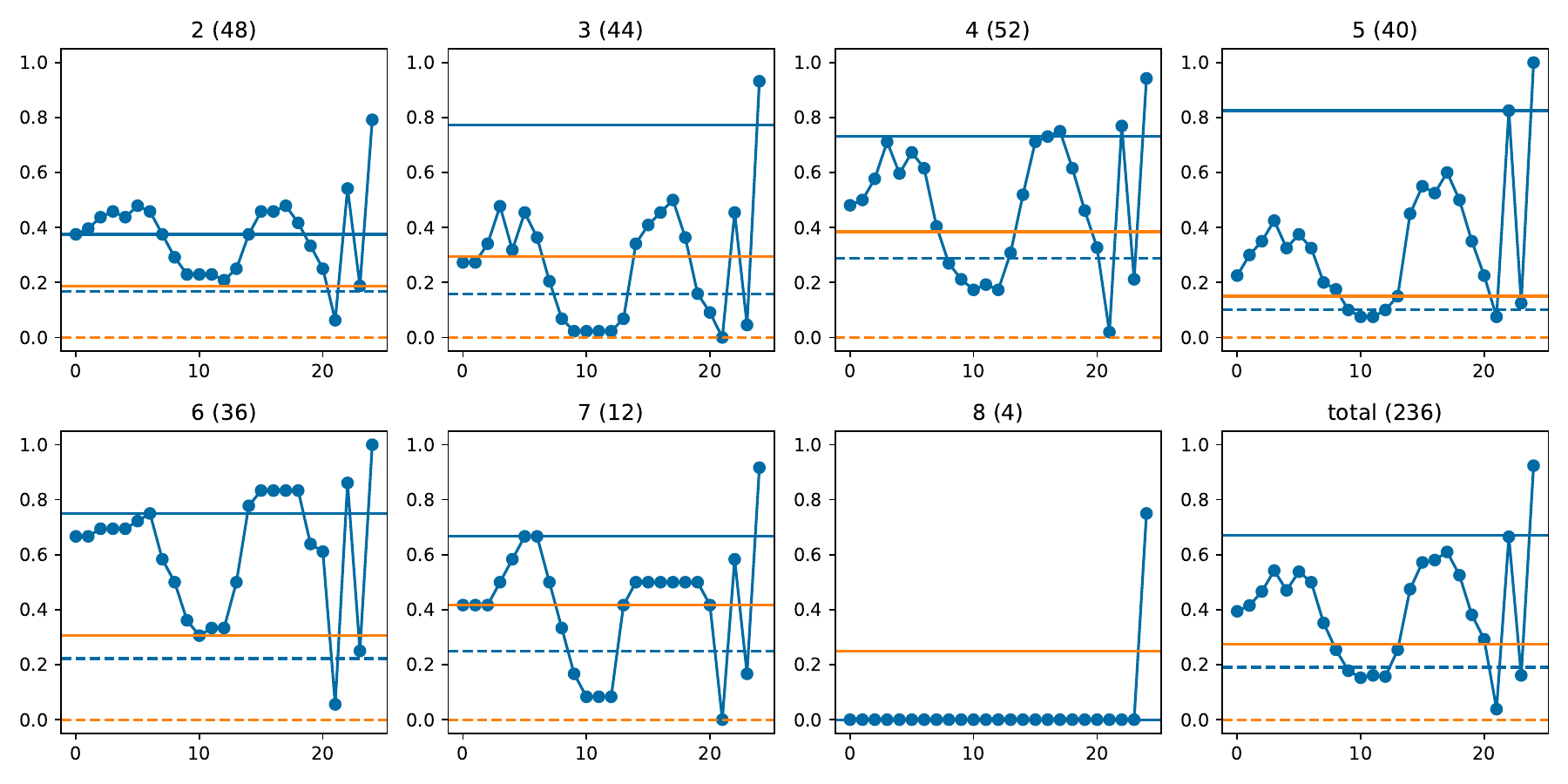}
  \includegraphics[width=\linewidth]{pdfs/model-comparison-legend.pdf}
  \caption{PanPhon feature distance-wise success rates on TIMIT using WavLM representations.}
  \label{fig:pfer-timit}
  \includegraphics[width=\linewidth]{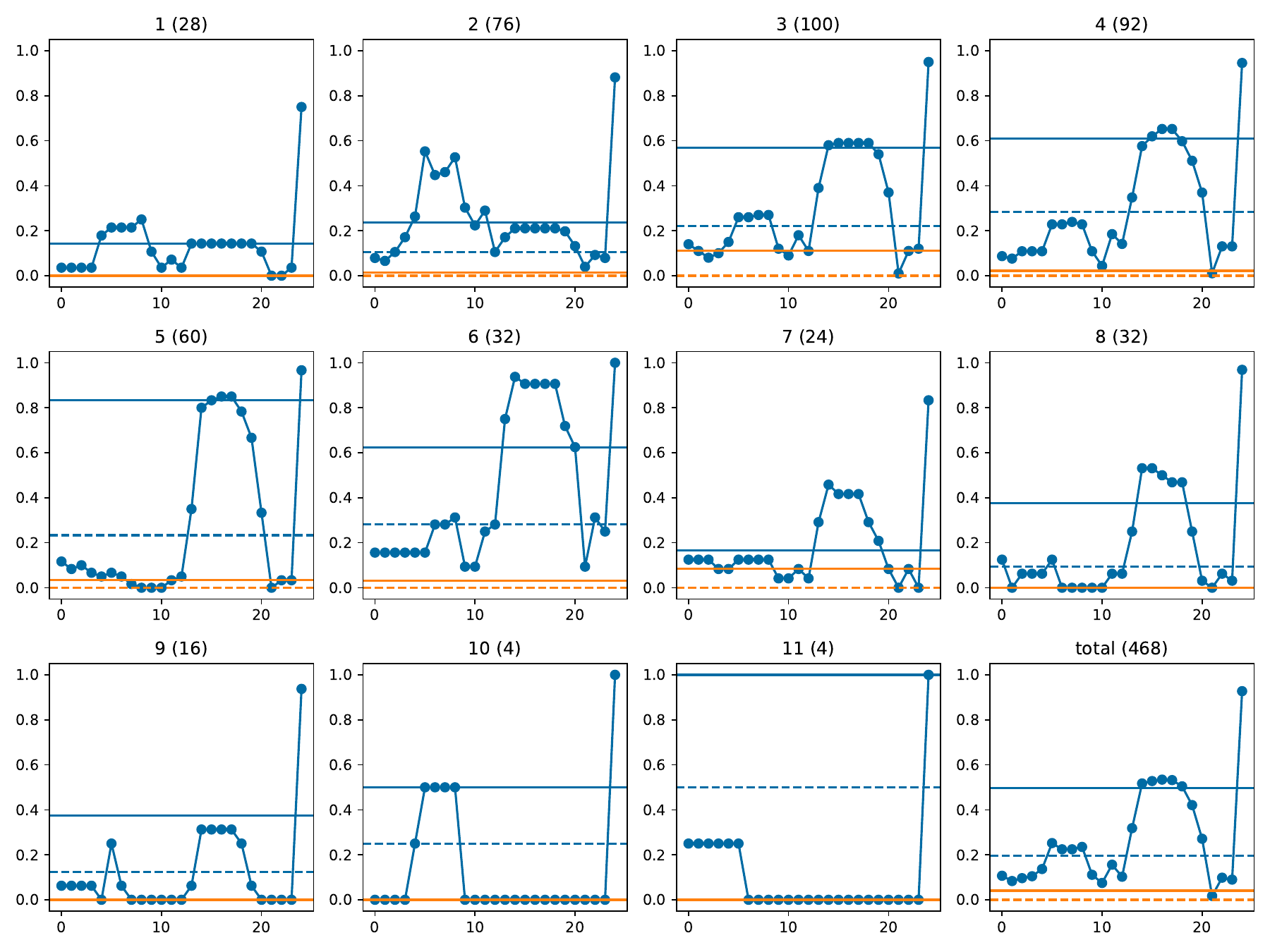}
  \includegraphics[width=\linewidth]{pdfs/model-comparison-legend.pdf}
  \caption{PanPhon feature distance-wise success rates on VoxAngeles using WavLM representations.}
  \label{fig:pfer-voxangeles}
\end{figure}

\subsubsection{Results}
For both TIMIT (\Cref{fig:pfer-timit}) and VoxAngeles (\Cref{fig:pfer-voxangeles}), the layerwise trends remain largely consistent across different phonological distances, with no variation observed as distance increases.
One exception occurs in distance bins containing only four quadruplets.
However, these cases arise from permutations of a single phone set, similar to \cref{ss:layerwise-panphon}.

\subsection{Measuring sample efficiency for phonological vector extraction}\label{ss:phono-samplecount}
\begin{figure*}[t]
\centering
  \includegraphics[width=\linewidth]{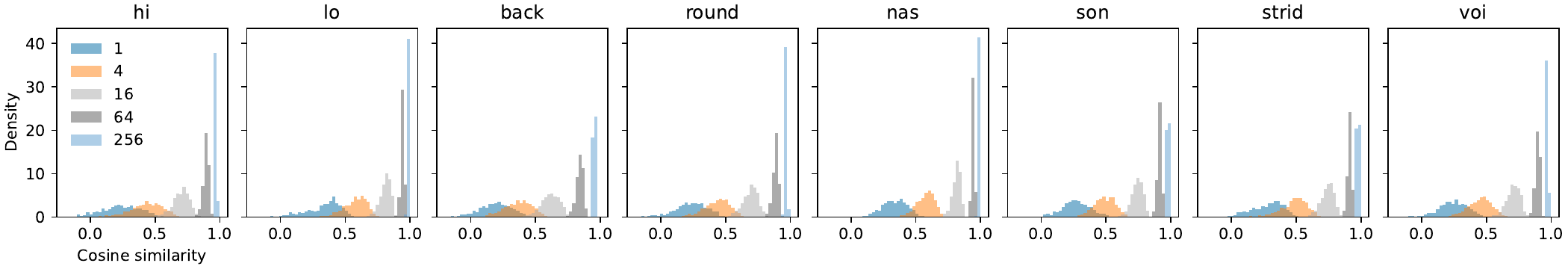}
  \includegraphics[width=\linewidth]{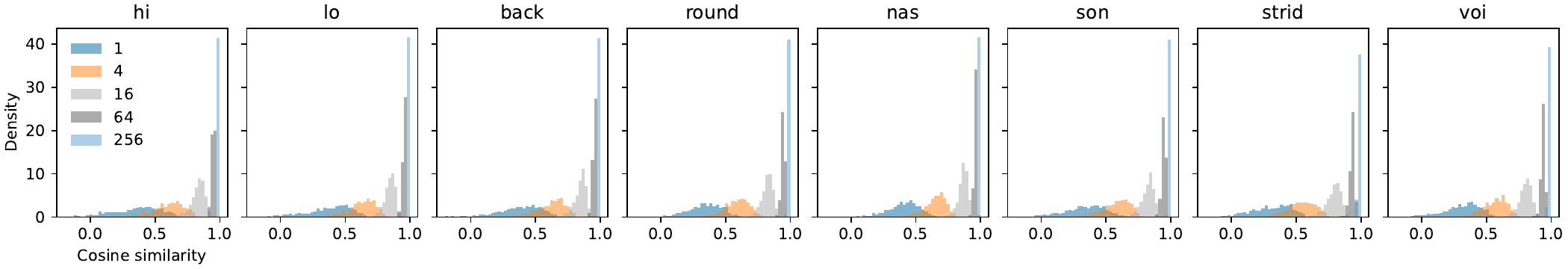}
  \caption{Approximating various phonological vectors on TIMIT (upper) and VoxAngeles (lower) with different sample size $N=\{1, 4, 16, 64, 256\}$.
  Each histogram depicts the distribution of cosine similarities.
  }
  \label{fig:phonosample}
\end{figure*}

\subsubsection{Settings}
In this experiment, we empirically evaluate how many phone representations are needed to extract accurate phonological vectors.
To measure the accuracy of extracted phonological vectors, we treat the vector extracted from the full training dataset in \cref{ss:modify} as the ground truth.
We then approximate this vector using smaller subsets of the training data.
We quantify approximation quality using cosine similarity.

We evaluate eight phonological vectors in \cref{sss:synth-phonovectors}.
For each positive and negative averaged representation in \Cref{eq:phonovec}, we randomly sample $N$ representations with replacement.
We repeat the approximation 1000 times with random samples and visualize the resulting cosine similarities using histograms.
We evaluate $N\in\{1, 4, 16, 64, 256\}$ sampled representations.

\subsubsection{Results}
For both TIMIT and VoxAngeles (\Cref{fig:phonosample}), cosine similarity consistently improves as $N$ increases across all phonological vectors.
When $N=256$, cosine similarity approaches 1.0, suggesting that a few hundred samples suffice for accurate approximation.

\subsection{Using a single phone pair for phonological vectors}\label{ss:phono-singlepair}
\begin{figure*}[t]
\centering
  \includegraphics[width=\linewidth]{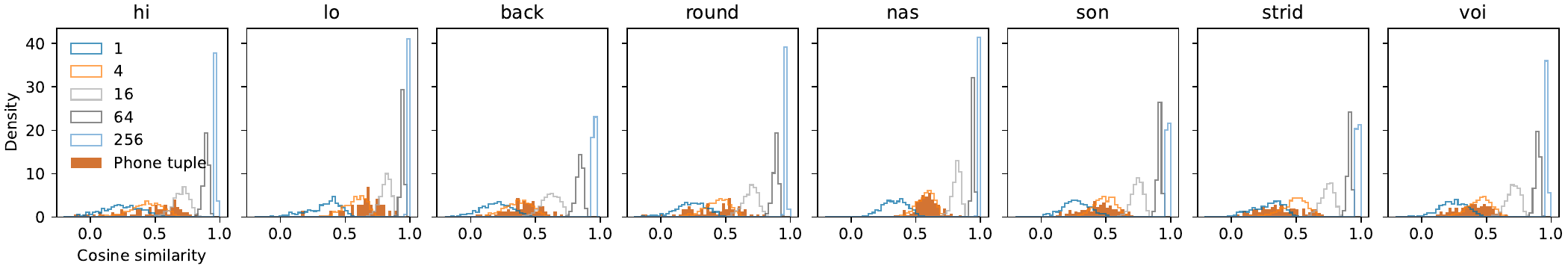}
  \includegraphics[width=\linewidth]{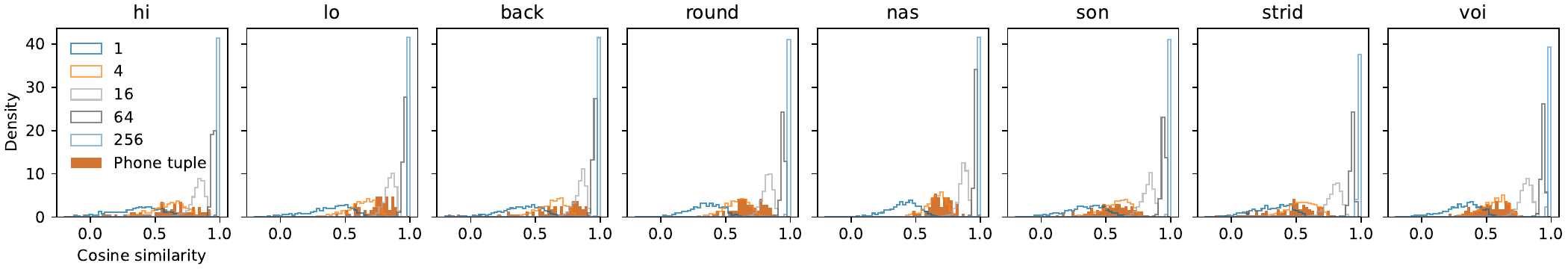}
  \caption{
  Approximating various phonological vectors on TIMIT (upper) and VoxAngeles (lower) using a fixed phone pair.
  Each histogram depicts the distribution of cosine similarities.
  \Cref{fig:phonosample} is overlaid for comparison.
  }
  \label{fig:phonophone}
\end{figure*}

\subsubsection{Settings}
We compare phonological vectors extracted using the full dataset with those obtained using only a single phone pair for the positive and negative averaged representations in \cref{ss:modify}.
For example, for the voicing phonological vector, we evaluate whether using only the ([b], [p]) pair leads to a good approximation.

We use the results from \cref{ss:phono-samplecount} as baselines for comparison.
Similarly, the vector obtained from the full training dataset is treated as the ground truth.
For each phonological vector, we evaluate all possible phone pairs and report the resulting distribution of cosine similarities with the ground truth via histograms.

\subsubsection{Results}
\Cref{fig:phonophone} shows the distribution of cosine similarities, using \cref{ss:phono-samplecount} as the baseline.
In TIMIT, each phone has few hundred to more than a thousand samples.
However, the cosine similarity remains around $0.5$, similar to using only four samples in the subsampling experiment.
These results indicate that using a fixed phone pair produces different vectors from those obtained using diverse phones, despite maintaining positive cosine similarity with the ground truth.
This is likely because the two phones in a pair often differ in multiple phonological features, causing the resulting vector to capture mixed contrasts.

\begin{figure}[t]
\centering
  \includegraphics[width=\linewidth]{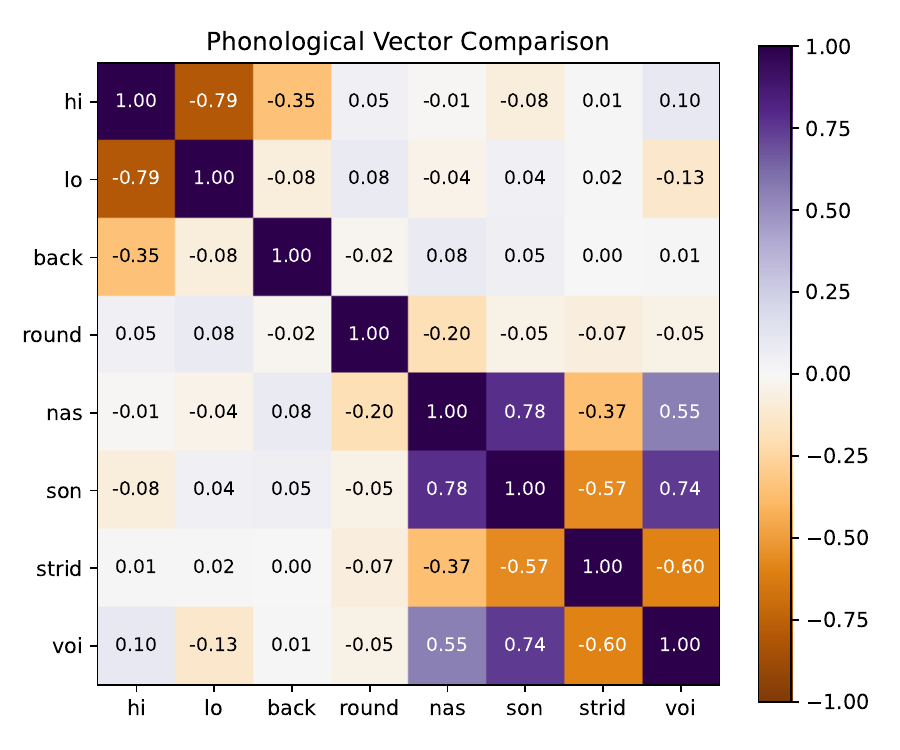}
  \caption{
  Cosine similarities between different phonological vectors drawn from TIMIT.
  }
  \label{fig:phonomatrix-timit}
  \includegraphics[width=\linewidth]{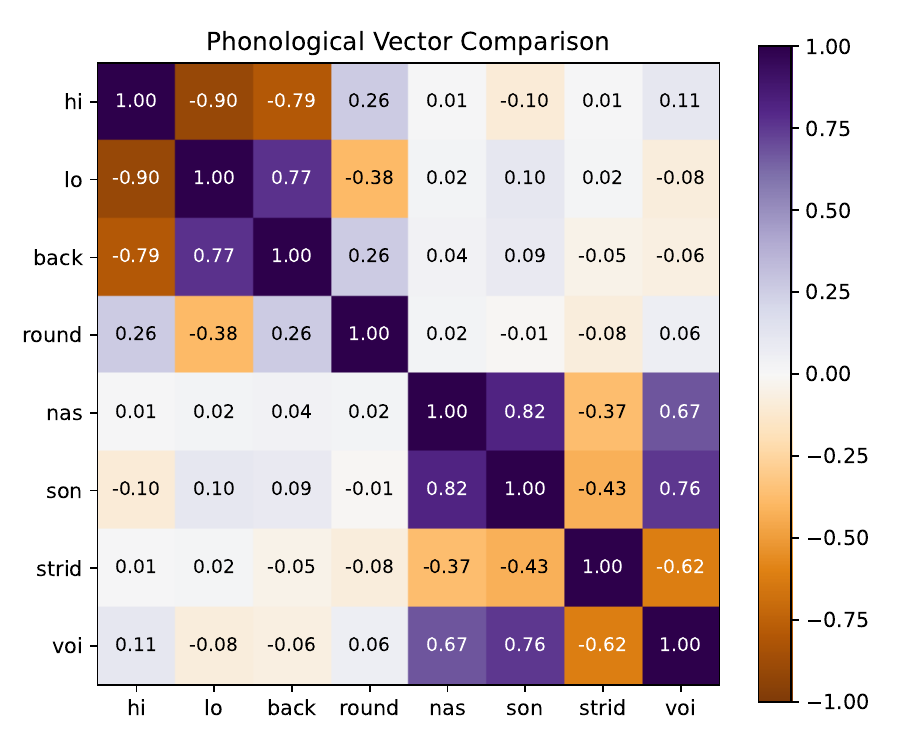}
  \caption{
  Cosine similarities between different phonological vectors drawn from VoxAngeles.
  }
  \label{fig:phonomatrix-voxangeles}
\end{figure}

\subsection{Comparing similarities between phonological vectors}\label{ss:phono-compare}
\subsubsection{Settings}
We investigate the relationships among different phonological vectors by measuring their pairwise cosine similarities.
Specifically, we compute cosine similarities between the eight phonological vectors defined in \cref{sss:synth-phonovectors}.

\subsubsection{Results}
\Cref{fig:phonomatrix-timit,fig:phonomatrix-voxangeles} visualize the pairwise cosine similarities between phonological vectors for TIMIT and VoxAngeles.
We mainly discuss cases with absolute cosine similarity values that are greater than $0.5$.
Across both datasets, we observe several phonologically interpretable patterns.
First of all, vowel-related and consonant-related phonological vectors exhibit near-orthogonal similarities.

For vowels, the high and low vectors show strong negative cosine similarity, reflecting their opposing acoustic properties.
Additionally, for VoxAngeles, round vector show positive similarity with high and back, and negative with low.
It is acoustically consistent with the rounding, which lowers both formants, F1 and F2.
High vowels are characterized by low F1, low vowels by higher F1, and back vowels by lower F2.

For consonants, nasal, sonorant, and voice vectors exhibit positive similarity, where nasals are always sonorants, and sonorants are almost always voiced.
Also, strident and sonorant vectors show negative similarity, where stridents are not sonorant and vice versa.

Overall, these results demonstrate that cosine similarities between extracted phonological vectors capture meaningful phonological relationships.

\subsection{Resynthesis works for unseen languages}\label{ss:voxangeles-synth}
\subsubsection{Settings}
Following \cref{sss:scatter-analysis}, we use the same settings of \cref{ss:exp2-method} by using representations from the final layer of WavLM.
As mentioned in \cref{ss:synth-corr-measure}, we construct train-test splits by randomly selecting languages for evaluation.
For resynthesis, as described in \cref{ss:vocos}, we use the vocoder trained on FLEURS-R, a multilingual speech dataset.

\subsubsection{Results}
A comparison between \Cref{fig:scatter-voxangeles} and \Cref{fig:scatter-timit} shows that VoxAngeles with FLEURS-R-trained vocoder exhibits trends that are nearly identical to those observed on TIMIT with LibriTTS-trained vocoder.
These include monotonic relationships between acoustic measurements and the phonological vector scale with the smooth interpolation and extrapolation behavior.
The close correspondence between seen and unseen language settings indicates that WavLM representations generalizes effectively to unseen languages through phonological vectors.

\begin{figure}[t]
  \centering
  \includegraphics[width=\linewidth]{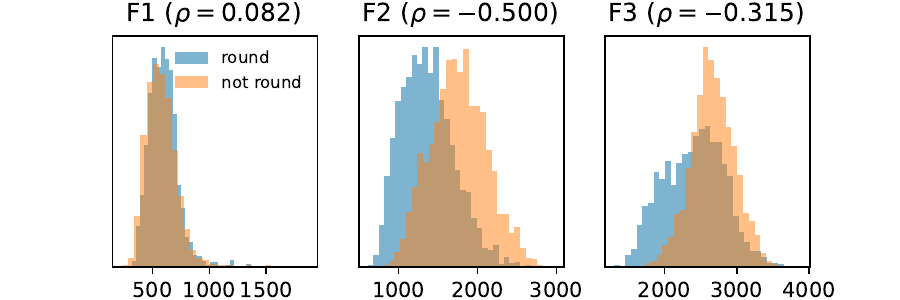}
  \includegraphics[width=\linewidth]{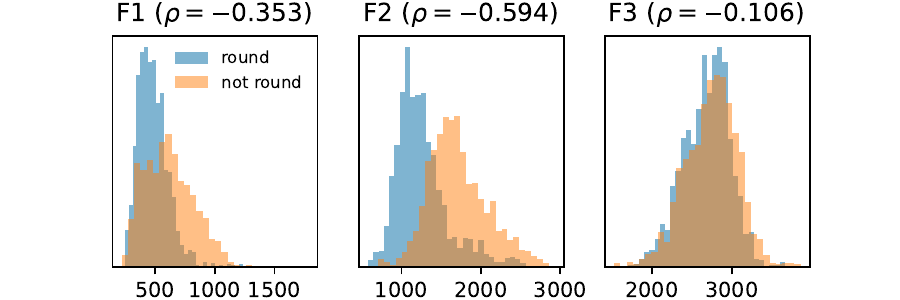}
  \caption{
    Acoustic measurements of F1, F2, and F3 for rounded (blue) and unrounded (orange) vowels on TIMIT (upper) and VoxAngeles (lower).
    $\rho$ indicates Spearman's rank correlation coefficient between the roundness and the acoustic measurements.
  }
  \label{fig:density-round}
\end{figure}

\begin{figure}[t]
  \centering
  \includegraphics[width=\linewidth]{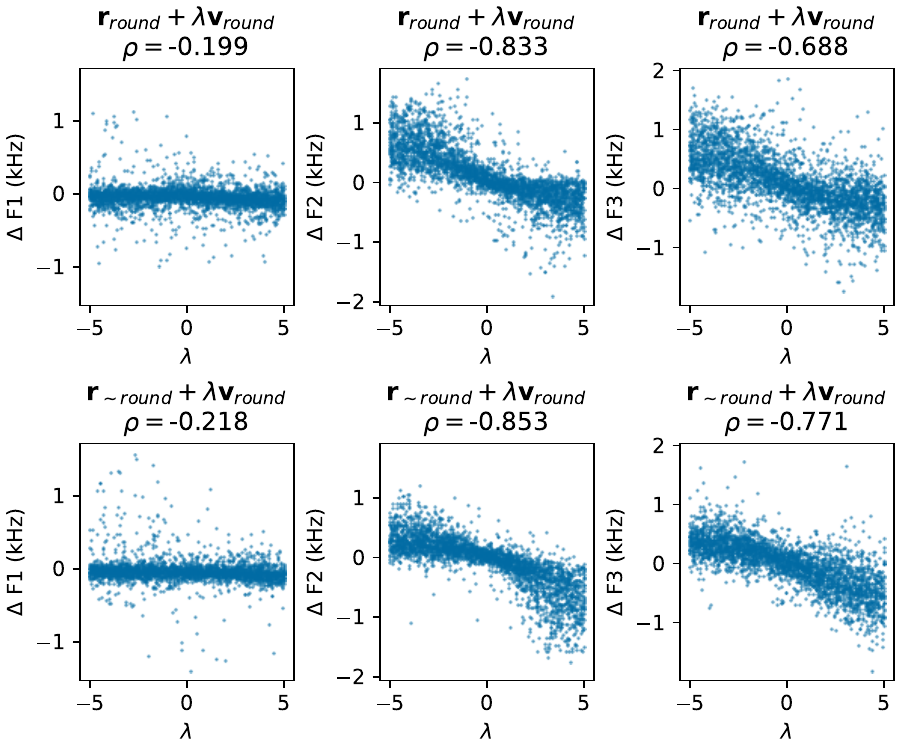}
  \includegraphics[width=\linewidth]{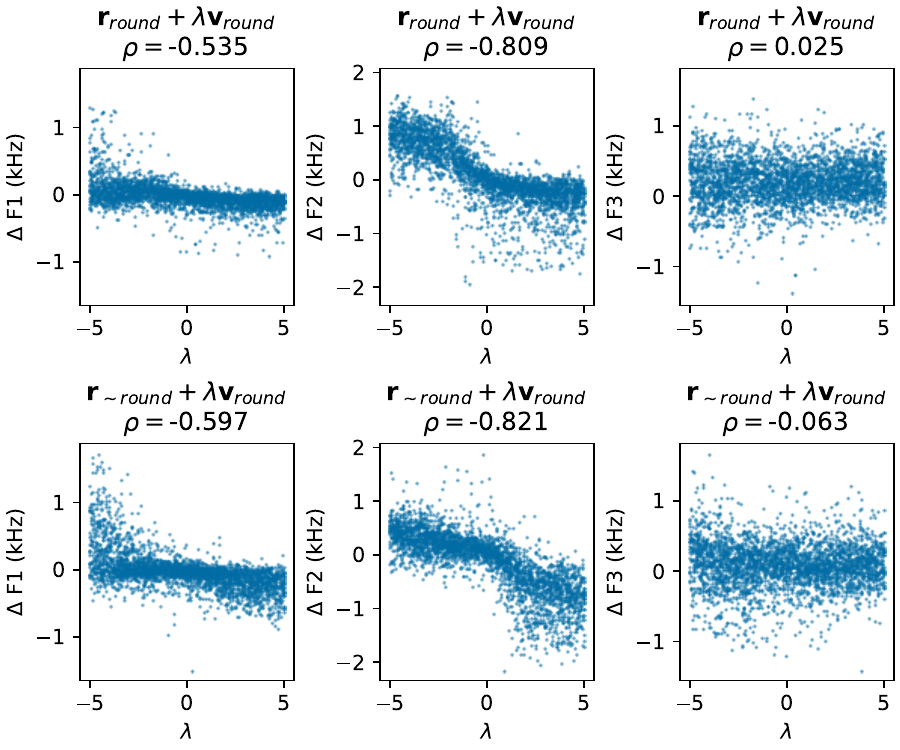}
  \caption{
    Comparing the phonological vector weight $\lambda$ with acoustic measurements on TIMIT (upper) VoxAngeles (lower) using WavLM.
    We observe three measurements, F1, F2, and F3, for the round vector.
    $\rho$ indicates Spearman's rank correlation coefficient.
  }
  \label{fig:scatter-round}
\end{figure}

\subsection{Case study: vowel rounding}\label{ss:synth-rounding}
\subsubsection{Settings}
To examine how a phonological vector influences multiple acoustic measurements, we analyze the first three formants (F1, F2, F3) for the rounding vector.
Lip rounding lengthens the front cavity of the vocal tract, which generally lowers all formants, while the specific behavior of each formant varies by language \cite{ladefoged1996elements,ladefoged2012vowels}.

To compare the behavior of the phonological vector with actual acoustic measurements, we first measure the formant distributions of rounded and unrounded vowels in the datasets.
We then examine whether the strength of the phonological vector $\lambda$ reflects these dataset-specific patterns.

\subsubsection{Results}
We first show the histograms of formant measurements of the vowels from TIMIT and VoxAngeles in \Cref{fig:density-round}.
For TIMIT, F2 and F3 differences are more prominent than F1, whereas for VoxAngeles the differences are more pronounced in F1 and F2 than in F3.

We then examine how formants vary with the strength of the phonological vector $\lambda$ in \Cref{fig:scatter-round}.
The same dataset-specific patterns emerge: the weakest correlation appears in F1 for TIMIT and in F3 for VoxAngeles.
This suggests that the rounding phonological vector captures dataset-specific phonetic realizations in a data-driven manner.

\begin{figure*}[t]
\centering
  \includegraphics[width=\linewidth]{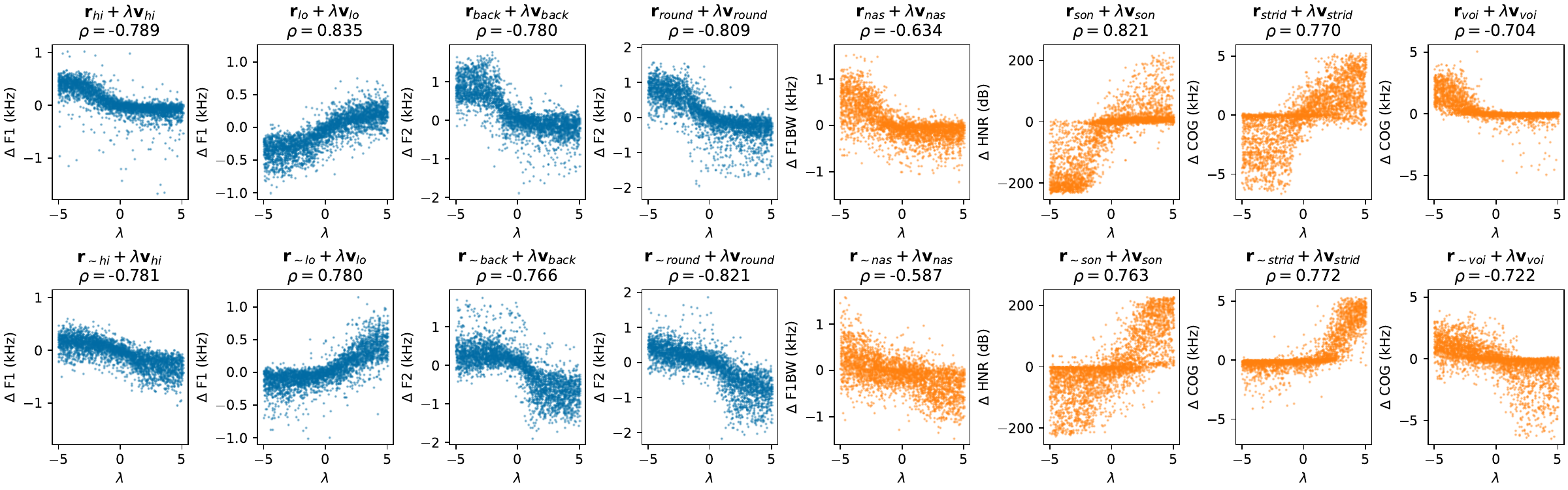}
  \caption{
    Comparing the phonological vector weight $\lambda$ with acoustic measurements on VoxAngeles using WavLM.
    $\rho$ indicates Spearman's rank correlation coefficient.
    Blue and orange plots indicate vowels and consonants, respectively.
  }
  \label{fig:scatter-voxangeles}
\end{figure*}

\subsection{Acoustic measurements remain stable after resynthesis}\label{ss:vocos-stability}
\subsubsection{Settings}
To assess the stability of the trained vocoder and the acoustic measurements, we compare the acoustic measurements computed from the original waveform $\mathbf{x}$ and from the resynthesized speech $\tilde{\mathbf{x}} = f^{-1}(f(\mathbf{x}))$.
We set $\lambda = 0$, corresponding to identity resynthesis without modifying the WavLM representations.

\begin{figure*}[t]
\centering
  \includegraphics[width=\linewidth]{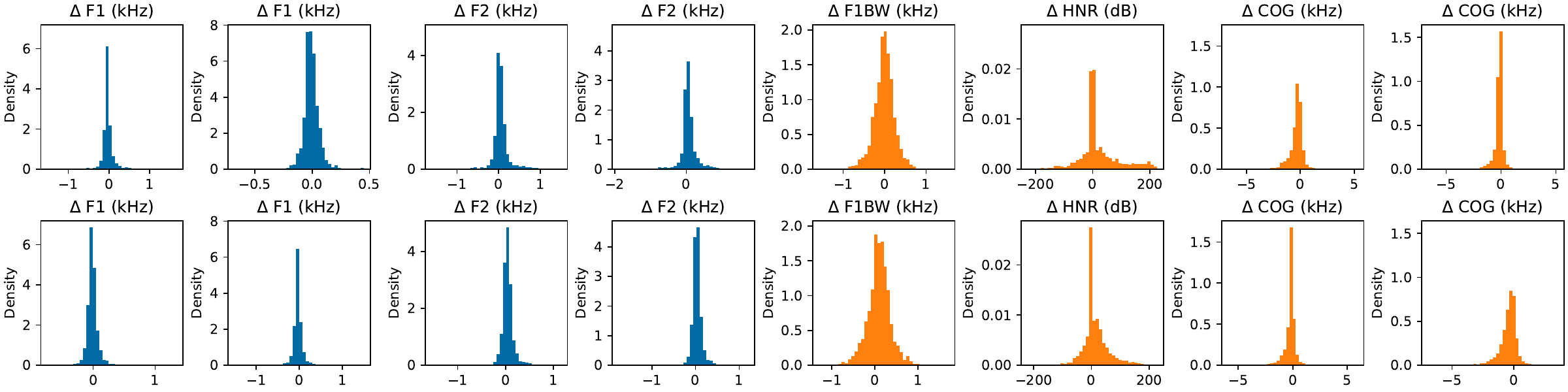}
  \includegraphics[width=\linewidth]{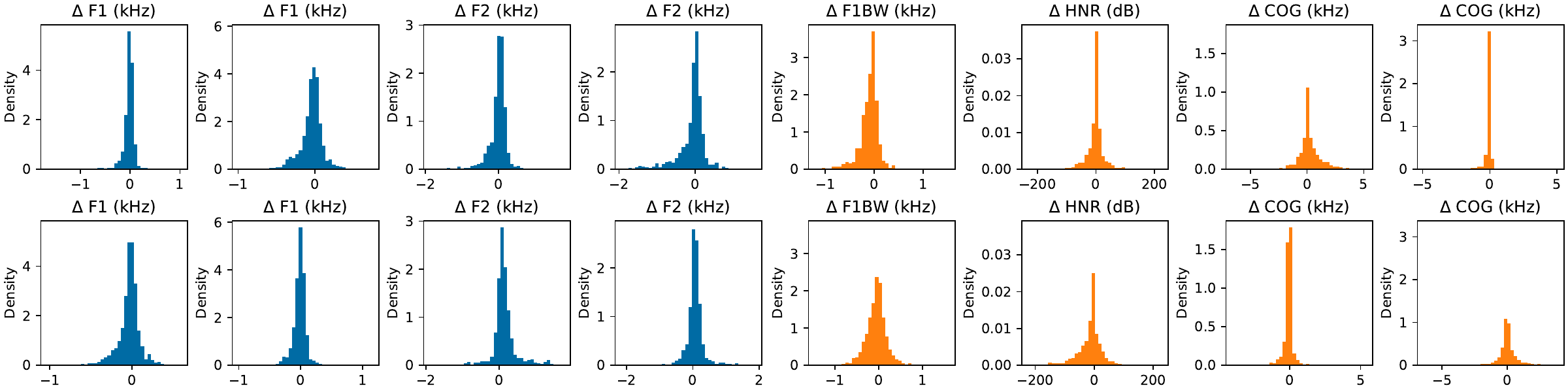}
  \caption{
    Comparing the acoustic measurements of original and synthesized speech ($\lambda=0$) on TIMIT (upper two rows) and VoxAngeles (lower two rows).
    We use the same range of x-axis from y-axis in \Cref{fig:scatter-timit,fig:scatter-voxangeles}.
    We observe that the differences are highly centralized to zero, ensuring the stability of resynthesis through the vocoder.
  }
  \label{fig:density-error}
\end{figure*}

\subsubsection{Results}
We visualize the differences in acoustic measurements using density plots in \Cref{fig:density-error}.
Across all tested consonants and vowels, the distributions are tightly concentrated around zero, indicating that the resynthesis and acoustic measurement pipeline leads to reliable analyses.

\subsection{Phonological vectors from MFCCs are ineffective for resynthesis}\label{ss:mfcc-synth}

\subsubsection{Settings}
Following \cref{sss:scatter-analysis} and \cref{ss:voxangeles-synth}, we adopt the same settings of \cref{ss:exp2-method}.
We evaluate both audio-sliced and feature-sliced MFCC representations, as \cref{ss:feat.vs.audio} demonstrated that audio slicing can improve the effectiveness of MFCCs.

\begin{figure*}[t]
\centering
  \includegraphics[width=\linewidth]{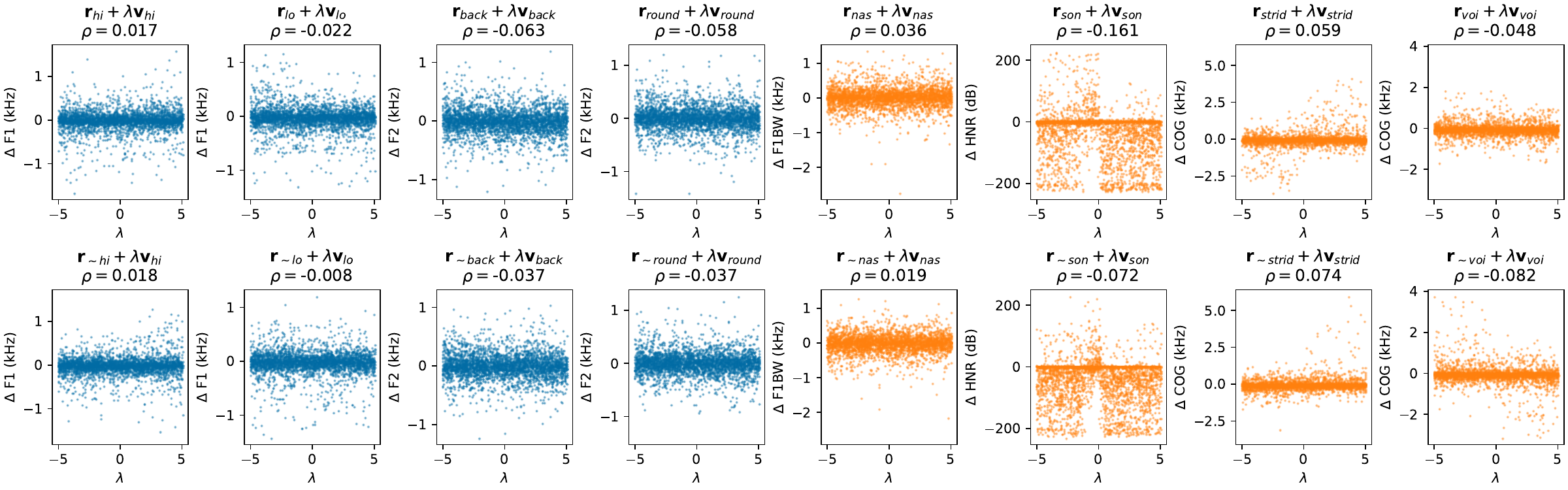}
  \includegraphics[width=\linewidth]{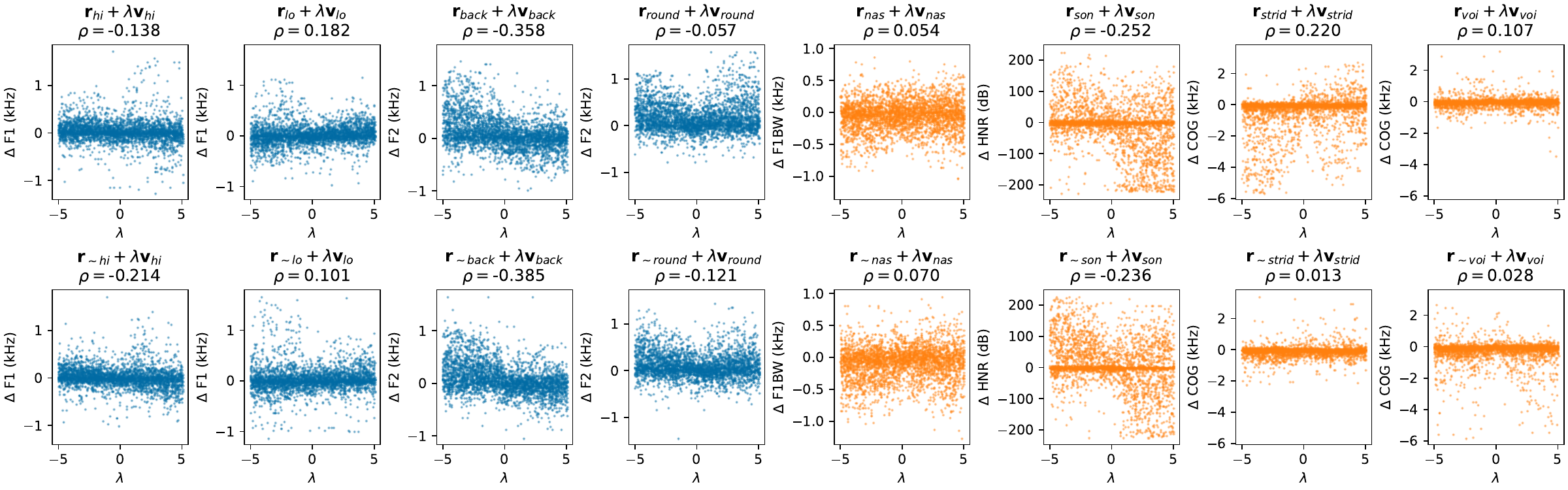}
  \caption{
    Comparing the phonological vector weight $\lambda$ with acoustic measurements on TIMIT (upper two rows) and VoxAngeles (lower two rows) using phonological vectors from audio-sliced MFCC.
    $\rho$ indicates Spearman's rank correlation coefficient.
    Blue and orange plots indicate vowels and consonants, respectively.
    There is little to no controllability, with the exception of weak correlation on back vowel on VoxAngeles.
  }
  \label{fig:scatter-mfcc}
  \includegraphics[width=\linewidth]{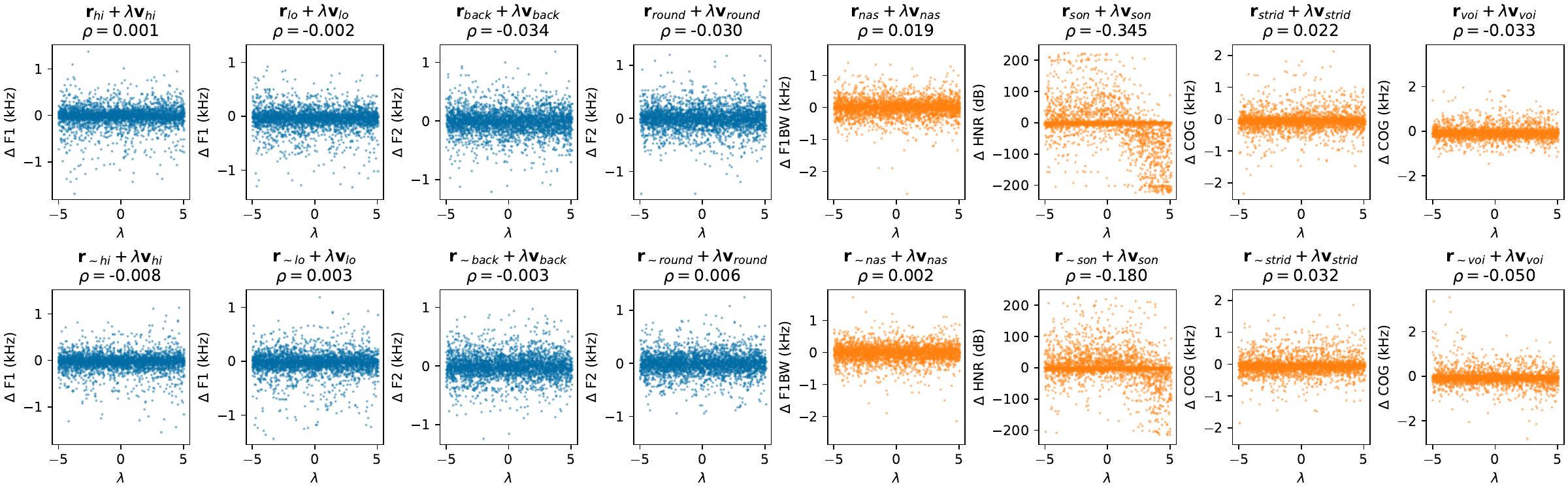}
  \includegraphics[width=\linewidth]{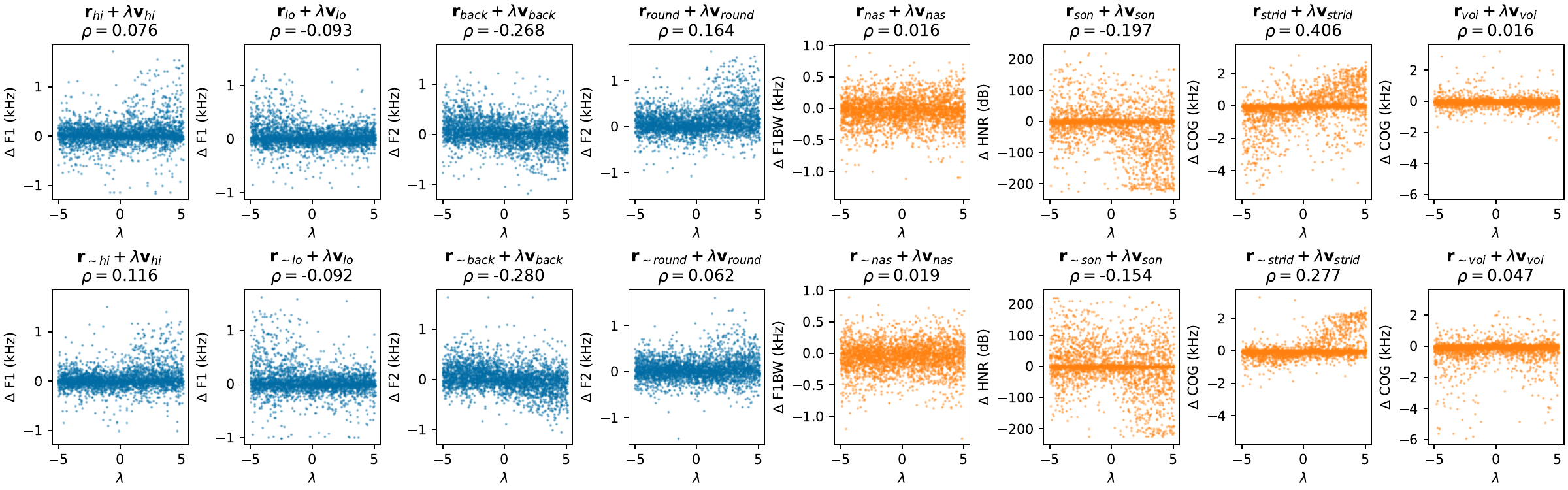}
  \caption{
    Comparing the phonological vector weight $\lambda$ with acoustic measurements on TIMIT (upper two rows) and VoxAngeles (lower two rows) using phonological vectors from feature-sliced MFCC.
    $\rho$ indicates Spearman's rank correlation coefficient.
    Blue and orange plots indicate vowels and consonants, respectively.
    Similar to \Cref{fig:scatter-mfcc}, there is little to no controllability.
  }
  \label{fig:scatter-mfcc-feat}
\end{figure*}

\subsubsection{Results}
\Cref{fig:scatter-mfcc,fig:scatter-mfcc-feat} show the results for audio-sliced and feature-sliced MFCCs, respectively.
Across both slicing strategies, the acoustic measurements exhibit little to no correlation with the scale of the corresponding phonological vectors.
The only exceptions are weak correlations observed for back vowels by the audio-sliced MFCCs and stridents by the feature-sliced MFCCs.
However, these effects are inconsistent, particularly given that no corresponding correlations are observed on TIMIT. 
Moreover, observed correlations are substantially waker than that of WavLM, as shown in \Cref{fig:scatter-timit,fig:scatter-voxangeles}.
Taken together, these results indicate that MFCC-derived representations are ineffective as phonological vectors for synthesis experiments.
\end{document}